\documentclass[12pt]{article}
\usepackage{amsfonts}
\usepackage{amssymb}
\usepackage{tikz}
\usepackage{amsmath,amssymb}
\usepackage{bbm}

\def\sfrac#1#2{{\textstyle\frac{#1}{#2}}}

\newcommand{\CT}{\mathcal{T}} 
\newcommand{\CO}{\mathcal{O}}    

\newcommand{\CH}{\mathcal{H}}

\newcommand{\Z}{\mathbb{Z}}   
   
\newcommand{\R}{\mathbb{R}}     
\newcommand{\C}{\mathbb{C}}     
\newcommand{\CPP}{{\mathbb{C}P}}    
\newcommand{\Nbb}{{\mathbb{N}}}    

\newcommand{\im}{\mathrm{i}} 
 
\newcommand{\dd}{\mathrm{d}}     
\newcommand{\dpar}{\partial}     
\newcommand{\hra}{{\hookrightarrow}}     
\newcommand{\diag}{{\mathrm{diag}}}     

\newcommand{\+}{\dagger}

\newcommand{\sU}{\mathrm{U}}     
\newcommand{\sSU}{\mathrm{SU}}     
     
\newcommand{\sSO}{\mathrm{SO}}     
\newcommand{\sGL}{\mathrm{GL}}  
 
\newcommand{\ru}{\mathrm{u}}

\newcommand{\al}{{{\alpha}}} 
  
\newcommand{\dal}{{{\dot\alpha}}}

\newcommand{\vph}{{{\varphi}}}

\newcommand{\zb}{{\bar{z}}}
\newcommand{\zd}{{\dot{z}}}

\newcommand{\Av}{{A_{\sf{vac}}}}
\newcommand{\Lv}{{L_{\sf{v}}}}

\newcommand{\qv}{{q_{\sf{v}}}}
\newcommand{\Fv}{{F_{\sf{vac}}}}

\newcommand{\und}{{\quad\mathrm{and}\quad}}
\newcommand{\with}{{\quad\mathrm{with}\quad}}
\newcommand{\for}{{\quad \mathrm{for}\quad }}

\makeatletter

\@addtoreset{equation}{section}
\makeatother

\begin{document}
\begin{titlepage}
\setcounter{page}{0}
.
\vskip 20mm
\begin{center}
{\large \bf The Geometry Underlying the Quantum Harmonic Oscillator}\\
\vskip 1.2cm
{\large Alexander D. Popov}
\vskip 0.7cm
{\em Institut f\"{u}r Theoretische Physik,
Leibniz Universit\"{a}t Hannover\\
Appelstra{\ss}e 2, 30167 Hannover, Germany}\\
{Email: alexander.popov@itp.uni-hannover.de}
\vskip 1cm
\end{center}
\begin{center}
{\bf Abstract}
\end{center}
We consider two-dimensional harmonic oscillator in the complex Bargmann-Fock-Segal representation with $T^*{\mathbb R}^{2}={\mathbb C}^2$ as classical phase space. We show that the eigenfunctions $\psi_n$ of the quantum Hamiltonian correspond to complex radial coordinates in the reduced phase space ${\mathbb C}^2/{\mathbb Z}_n\subset{\mathbb C}^2$. They describe ${\mathbb Z}_n$-invariant motion of particle along a circle $S^1$ in lens space $S^3/{\mathbb Z}_n\subset{\mathbb C}^2/{\mathbb Z}_n$, where ${\mathbb Z}_n$ is the cyclic group of rotation by an angle $2\pi/n$ on the circle $S^1$, $n=1,2,...\,$. Thus the general solution  of the Schr\"odinger equation carries information about an infinite number of admissible  classical states $\psi_n$ that can be mapped to other states after lifting into the quantum bundle. We show that in the Kepler/hydrogen atom problem there is a similar correspondence between classical and quantum states.

\end{titlepage}

\newpage
\setcounter{page}{1}

\section{Introduction and summary}

\noindent   The quantum harmonic oscillator is one of the most important models of quantum theory. It appears in almost every area of physics, since any system near the minimum of its potential energy is approximately described by a harmonic oscillator. This is an integrable model that introduces creation and annihilation operator, a ground state, a zero-point energy, and a Fock space, i.e. everything needed to formulate quantum field theory \cite{Dirac}. It is believed that the model is so simple that everything about it is known at both the classical and quantum level. We will show that this is not the case using the example of a harmonic oscillator in two dimensions. For this we will translate all the concepts associated with the oscillator (ground state, creation/annihilation operators, etc.) into the language of geometry and show that algebraic geometry dictates a different picture of the correspondence between the quantum and classical states of the oscillator than is commonly believed. Namely, passing from wave functions to coordinates of quantum bundles, we associate with each basis vector (except the ground state) of its Hilbert space a certain submanifold in the classical phase space, defined by a finite cyclic group. In this case, the transition from classical to quantum states corresponds to the transition from the coordinates of the bundle to its sections.

We will consider the harmonic oscillator in the complex Bargmann-Fock-Segal representation, when dimensionless complex coordinates $z_\al$ are specified on the phase space $T^*\R^2\cong\C^2$ of the classical oscillator, $\al =0,1$ \cite{Dirac, Bar, Segal, Hall}. The quantum bundle in this case is defined as a complex space 
\begin{equation}\label{1.1}
\Lv = \C^2\times\C\ni (z_0, z_1, \Psi )\ ,
\end{equation}
where $\Psi$ is a complex coordinate on the fibre $\C$ over the phase space $\C^2$ (see e.g. \cite{Sni, Wood, Hurt}). A classical oscillator is described by a point $z_\al(t)$ moving in phase space $\C^2$, where $t\in\R$ is the time parameter. The quantum oscillator is described by holomorphic sections $\Psi(z_0, z_1, t)$ of the bundle \eqref{1.1}, that is, the surface $(z_0, z_1, \Psi(z_0, z_1))\subset\Lv$ is considered instead of the point $(z_0, z_1)\in\C^2$. In this paper we consider an almost quantum oscillator as a point moving in the extended phase space \eqref{1.1} with three independent coordinates. There is no discussion of this intermediate case between the classical and quantum oscillator in the literature. We will show that it is at this level we can better understand the mathematical and physical nature of the wave function. In particular, we will show that the ground state is a motion in the fibre $\C$ of the bundle \eqref{1.1} over the phase space $\C^2$, i.e. outside the classical phase space, which is a key to the transition from the classical to the quantum description.

The eigenfunctions of the Hamiltonian of a quantum harmonic oscillator are 
\begin{equation}\label{1.2}
\Psi_n(z)=\psi_n(z)\psi_0^{\sf{v}}(z)\ , \quad \psi_n(z)=\sum_{m=0}^n b_{nm}\psi_{nm}(z)\ ,\quad
\psi_{nm}(z)=\frac{z_0^{n-m}z_1^m}{\sqrt{(n-m)!m!}}\ ,
\end{equation}
where $\psi_0^{\sf{v}}(z)$ is the ground state, and $\psi_{nm}(z)\psi_0^{\sf{v}}(z)$ form a basis of the Hilbert space $\CH$ of the quantum oscillator, $n=0,1,...\,$. The energy associated with the function $\psi_n(z)$ is $E_n=\hbar\omega n$, and the ground state $\Psi_0=b_{00}^{}\psi_0^{\sf{v}}=\psi_0^{\sf{v}}$ is associated with the energy $E_0^{\sf{v}}=\hbar\omega$. Note that the ground state $\psi_0^{\sf{v}}$ is included as a common factor in all basis vectors \eqref{1.2} of the Hilbert space $\CH$. Therefore, when considering the action of the structure group $\sU(1)_{\sf{v}}\subset\sGL(1,\C)_{\sf{v}}$ of the bundle \eqref{1.1} on sections, we assign weight 0 to the functions  $\psi_n(z)$ and weight 1 to the function $\psi_0^{\sf{v}}(z)$, i.e.  $\sU(1)_{\sf{v}}$ does not act on $\psi_n(z)$.

It is well known that homogeneous polynomials $\psi_n(z)$ are global sections of holomorphic bundles $\CO(n)$ over the Riemann sphere $\CPP^1$ with homogeneous coordinates $[z_0:z_1]$ (see e.g. \cite{Wood, Hurt, Wells}). As stated earlier, we want to consider at first an almost quantum oscillator, that is, a point moving in the extended phase space \eqref{1.1}. In this case, $\psi_n$ specifies a {\it coordinate} on the fibre bindle $\CO(n)$, not a section. Note that $\psi_{nm}$ are special cases of coordinates on the fibres of the same bundle $\CO(n)$ with an additionally defined action of the group of non-zero complex numbers $\C^*=\sGL(1,\C)$ on the total space of the bundle $\CO(n)$ (equivariant structure of the weight $m$). After lifting from $\CPP^1$ to $\C^2$, these bundles become trivial bundles over $\C^2$ with an equivariant structure determined by the numbers $n$ and $m$. After passing to sections, these numbers will determine the grading of the Hilbert space $\CH$.

In the case of bundles $\CO(n)$ under consideration, it follows from Geometric Invariant Theory (see e.g. \cite{Mumford, Thomas}) that the total space $\CO(n)^{\times}:=\CO(n){\setminus}\{\psi_n=0\}$ of the bundle $\CO(n)$ without zero section is biholomorphic to the orbifold $(\C^2{\setminus}\{0\})/\Z_n$ obtained from the phase space $\C^2{\setminus}\{0\}$ by reduction under the action of the group $\Z_n$. Here $\Z_n=\Z/n\Z$ is the group of rotation of the circle $S^1$ in the sphere  $S^3\subset\C^2{\setminus}\{0\}$ by an angle multiple $2\pi/n$. Both spaces $\CO(n)^{\times}$ and $(\C^2{\setminus}\{0\})/\Z_n$ have lens space $S^3/\Z_n$ as a subspace. Thus we have a correspondence
\begin{equation}\label{1.3}
\begin{picture}(100,60)
\put(0.0,0.0){\makebox(0,0)[r]{$\CO(n)^{\times}$}}
\put(-25.0,45.0){\makebox(0,0)[r]{$(z_0, z_1, \psi_n)\in$}}
\put(55.0,0.0){\makebox(0,0)[c]{$\supset S^3/\Z_n\subset$}}
\put(100.0,0.0){\makebox(0,0)[l]{$(\C^2{\setminus}\{0\})/\Z_n$}}
\put(44.0,45.0){\makebox(0,0)[c]{$Y_n=(\C^2{\setminus}\{0\})\times\C_{(n)}$}}
\put(-5.0,25.0){\makebox(0,0)[c]{$\C^*$}}
\put(100.0,25.0){\makebox(0,0)[c]{$\C^*$}}
\put(10.0,30.0){\vector(-1,-1){18}}
\put(80.0,30.0){\vector(1,-1){18}}
\end{picture}
\\[5pt]
\end{equation}
where projections from the space $Y_n$ are realized by factorization by the group $\C^*$, and the coordinate $\psi_n$ in the fibre of the bundle $Y_n$ over $\C^2{\setminus}\{0\}$ belongs to the space $\C_{(n)}$ of one-dimensional holomorphic representation of group $\C^*$ of weight $n$ ($z_\al$ have weight one). 

To describe the ground state $\psi_0^{\sf{v}}$ with $n=0$ we introduce a blow-up $\tilde\C^2=\CO(-1)\cong\CPP^1\sqcup(\C^2{\setminus}\{0\})$ of the space $\C^2$ in the origin by replacing the point $\{0\}$ in $\C^2=\{0\}\sqcup(\C^2{\setminus}\{0\})$ with a sphere $\CPP^1$. The ground state $\psi_0^{\sf{v}}\in \C_{(0)}$ is described by the correspondence
\begin{equation}\label{1.4}
\begin{picture}(100,60)
\put(20.0,0.0){\makebox(0,0)[r]{$([z_0:z_1], \psi_0^{\sf{v}})\in\CPP^1{\times}\C_{(0)}$}}
\put(50.0,0.0){\makebox(0,0)[l]{$\{0\}{\times}\C_{(0)}\ni \psi_0^{\sf{v}}$}}
\put(40.0,45.0){\makebox(0,0)[c]{$\tilde\C^2\times\C_{(0)}$}}
\put(-10.0,10.0){\vector(1,1){25}}
\put(30.0,35.0){\vector(1,-1){25}}
\end{picture}
\\[10pt]
\end{equation}
where $\CPP^1$ is embedded in $\CO(-1){=}\tilde\C^2$ as zero section.
This shows that the almost quantum oscillator in this state is a point rotating as $\psi_0^{\sf{v}}\exp(-\im\omega t)$ in the fibres $\C_{(0)}$ of the quantum bundle \eqref{1.1} and resting in the point $\{0\}$ of the phase space $\C^2$. In principle, one can consider any point in space $\C^2$, but it is preferable to place the particle at a fixed point  $\{0\}\in \C^2$.   The coordinate $\psi_0^{\sf{v}}\in \C_{(0)}$ corresponds to the global holomorphic section of the trivial bundle $\CO(0)$ over the sphere $\CPP^1$ which projects to the point $\{0\}\in\C^2$. The energy of this motion is equal to $E_0^{\sf v}=\hbar\omega$, and no ``vibrations" are observed around point $\{0\}\in\C^2$.  This energy curves the total space $\Lv$ of the bundle \eqref{1.1}, leading to a constant curvature $\Fv$ on it, but does not curve the base $\C^2$ of this bundle. 

In contrast to the ground state rotation, the rotation in the fibres of the bundle $\CO(n)^{\times}$ according to the diagram \eqref{1.3} is the motion of the particle in the conical space $(\C^2{\setminus}\{0\})/\Z_n$, which is embedded as a subspace in the phase space $\C^2{\setminus}\{0\}$ of the particle. Thus, in accordance with the product $\Psi_n=\psi_n\psi_0^{\sf{v}}$ from \eqref{1.2}, the motion of an almost quantum oscillator is a motion along a circle $S^1$ in the torus $S^1_{\sf cl}\times S^1_{\sf qu}$, where $S^1_{\sf cl}$ is a circle in the lens space $S^3/\Z_n\subset (\C^2{\setminus}\{0\})/\Z_n$, given by the coordinate $\psi_n\exp(-\im\omega nt)$ in the diagram \eqref{1.3}, and $S^1_{\sf qu}$ is a circle in the fibre of the quantum bundle \eqref{1.1}, given by the coordinate $\psi_0^{\sf{v}}\exp(-\im\omega t)$ in diagram \eqref{1.4}. The lens space $S^3/\Z_n$ is a fibre bundle over $\CPP^1$,
\begin{equation}\label{1.5}
S^3/\Z_n\ \stackrel{S^1_{\sf cl}}{\longrightarrow}\ \CPP^1\ ,
\end{equation}
coinciding with the Hopf fibration \cite{Hopf, Seifert} for $n=1$ since $\Z_1{=}$Id (standard classical oscillator). The circle $S^1_{\sf cl}$ in fibres of the bundle \eqref{1.5} is shortened, $0\leq\vph<2\pi/n$, and the angular velocity on it is equal to $\omega_n=\omega n$. The smooth manifold \eqref{1.5} is a submanifold of isomorphic spaces $\CO(n)^{\times}$ and $(\C^2{\setminus}\{0\})/\Z_n$ in \eqref{1.3}.

The subspace of the phase space $\C^2{\setminus}\{0\}$ that is a representative of the orbifold $(\C^2{\setminus}\{0\})/\Z_n$ is the phase space of $\Z_n$-invariant classical oscillator with $n\in\Nbb$. The transition from an almost quantum to quantum oscillator is accomplished by replacing a point in the spaces \eqref{1.1}, \eqref{1.3}-\eqref{1.5} with a section of the bundle $\CO(n){\otimes}\CO(0)$, which replaces the rotation of the point with the rotation of the sphere $\CPP^1$ as a whole in these spaces. In this case, $\psi_n\bar\psi_n\exp(-|z|^2)$ defines stationary states smeared around $\CPP^1\times S^1$ for $n=0$ and around $S^3/\Z_n$ for $n\geq 1$. The main part of the paper contains a detailed description of all these correspondences, including mapping $\psi_n\mapsto\psi_{n\pm 1}$ that changes the angular frequency
 $\omega_n$ and the energy $E_n$ of states due to the interaction with the connection in the quantum bundle $\Lv$. In Conclusions, we will show that the motion of an electron in a hydrogen atom is described by diagrams similar to \eqref{1.3} and \eqref{1.4} with $S^3/\Z_n$ from 
\eqref{1.3} replaced by $(S^3\times S^2)/\Z_{n-1}$  with $n\geq 2$ and $S^2\times S^1$ from \eqref{1.4} replaced by $S^2\times S^2\times S^1$ for the ground state with $n=1$. The main idea of this paper is to explore the geometry underlying classical and quantum states.

\section{Classical oscillator}

\noindent {\bf Hamiltonian.} A classical isotropic harmonic oscillator in two dimensions $\R^2$ is a particle of mass $m$ rotating around a circle $S^1$ in phase space $T^*\R^2=\R^4\cong\C^2$ with  angular velocity $\omega$. On this phase space, parametrized by coordinates $x_\al$ and momenta $p_\al$, we introduce dimensionless complex coordinates
\begin{equation}\label{2.1}
z_\al =\rho_\al e^{\im\vph_\al}=\frac{1}{r_0}(x_\al - \frac{\im}{m\omega}p_\al)\with r_0^2=\frac{2\hbar}{m\omega}\ ,
\end{equation}
where $r_0$ is the length parameter \cite{Dirac}. In these coordinates, the Hamiltonian function of the oscillator has the form
\begin{equation}\label{2.2}
H=\frac{p^2}{2m}+\frac{m\omega^2 x^2}{2}=\hbar\omega|z|^2
\for |z|^2=\delta^{\al\dal}z_\al\zb_\dal =\rho_0^2+\rho_1^2=:\rho^2\ ,
\end{equation}
where the Planck constant $\hbar$ appears due to its presence in the definition \eqref{2.1}. Here $p^2=p_0^2+p_1^2$, $x^2=x_0^2+x_1^2$, the bar denotes complex conjugation and $\al, \dal =0,1.$

\smallskip

\noindent {\bf Metric and 2-form.} Next we will consider the space $\C^2$ with the point $\{0\}=(0,0)$ removed. This space can be regarded as a cone over 3-sphere $S^3$,
\begin{equation}\label{2.3}
\C^2{\setminus}\{0\}=C(S^3)
\end{equation}
with the metric 
\begin{equation}\label{2.4}
g_{\C^2}^{}=\delta^{\al\dal}\dd z_\al\dd \zb_\dal=\dd\rho^2+\rho^2\dd\Omega^2_{S^3}\ ,
\end{equation}
where the radial coordinate $\rho$ is defined in \eqref{2.2}. On the phase space $\C^2{\setminus}\{0\}$ we also introduce the symplectic 2-form
\begin{equation}\label{2.5}
\Omega^{}_{\C^2}=-\im\hbar\delta^{\al\dal}\dd z_\al\wedge\dd \zb_\dal =\dd\left[\sfrac{\im\hbar}2\delta^{\al\dal}(\zb_\dal\dd z_\al -z_\al\dd\zb_\dal)\right]\ .
\end{equation}
The function \eqref{2.2} defines a Hamiltonian vector field of the form
\begin{equation}\label{2.6}
V_H^{}=\Omega^{}_{\dal\al}\dpar_{\zb_\dal}H\dpar_{z_\al}+\Omega^{}_{\al\dal}\dpar_{z_\al}H\dpar_{\zb_\dal}=\im\omega z_\al\dpar_{z_\al}-\im\omega \zb_\dal\dpar_{\zb_\dal}\ ,
\end{equation}
where $\dpar_{z_\al}:=\dpar/\dpar z_\al$ and $\dpar_{\zb_\dal}=\dpar/\dpar{\zb_\dal}$.
The point $\{0\}$ in phase space is fixed for the group generated by $V_H$. It is associated with the ground state $\psi_0^{\sf{v}}$ and we defer its consideration until Section 7.

\smallskip

\noindent {\bf Hopf fibration.} The vector field \eqref{2.6} is the generator of the group $\sU(1)$ acting on the level surface $S^3\subset\C^2{\setminus}\{0\}$,
\begin{equation}\label{2.7}
S^3:\  \rho^2=|z|^2=\gamma^2\ .
\end{equation}
Here $\gamma$ is any positive number. When choosing $\gamma=1$, function \eqref{2.2} will be equal $H=\hbar\omega$. The orbits of the group $\sU(1)$ in $S^3$ are given by formulae
\begin{equation}\label{2.8}
z_\al (t)=e^{tV_H}z_\al =e^{\im\omega t}z_\al\quad\Rightarrow\quad \zd_\al(t)=\im\omega z_\al(t)\ ,
\end{equation}
where $z_\al =z_\al(0)$. The orbit space is parametrized by a 2-sphere $S^2$ in $S^3$ defined by equivalence relation $z_\al (t)\sim z_\al(0)$. Quotienting the sphere $S^3$ by the action \eqref{2.8} of the group $\sU(1)$ yields the Hopf fibration
\begin{equation}\label{2.9}
\pi:\ S^3\ \stackrel{S^1}{\longrightarrow}\ S^2\cong\CPP^1
\end{equation}
such that $S^3$ is a nontrivial principal $\sU(1)$-bundle $S^3=P(S^2, \sU(1))$ over the Riemann sphere $\CPP^1$ \cite{Hopf, Seifert}.

\smallskip

\noindent {\bf Bundle $\CO(-1)^\times$.} We will introduce two regions on $\CPP^1=\C\cup\{\infty\}$,
\begin{equation}\label{2.10}
U_0:\ z_0\ne 0, z:=\frac{z_1}{z_0}\und U_1:\ z_1\ne 0, w:=\frac{z_0}{z_1}\ ,
\end{equation}
and denote by $[z_0:z_1]$ the homogeneous coordinates on $\CPP^1=U_0\cup U_1$. Using local coordinates \eqref{2.10}, on the space $\C^2{\setminus}\{0\}$ one can introduce coordinates
\begin{equation}\label{2.11}
(\psi_{-1}z_\al)=[\psi_{-1}\rho e^{\im\vph}e^{-\im\chi/2}]\frac{(1,z)}{(1+z\zb)^{1/2}}=
[\psi_{-1}\rho e^{\im\vph}e^{\im\chi/2}]\frac{(w,1)}{(1+w\bar w)^{1/2}}\ ,
\end{equation}
where
\begin{equation}\label{2.12}
\vph :=\sfrac12(\vph_0+\vph_1)\ ,\quad \chi :=\vph_1-\vph_0\und z=\frac{\rho_1}{\rho_0}e^{\im\chi}=\frac1w\ .
\end{equation}
Here, $\psi_{-1}$ is a constant such that $\psi_{-1}=0$ parametrizes the point $\{0\}\in\C^2$, and on $\C^2{\setminus}\{0\}$ one can always choose $\psi_{-1}=\gamma^{-1}$, so that condition \eqref{2.7} for \eqref{2.11} will define a unit sphere for $\tilde z_\al =\psi_{-1}z_\al$.

From \eqref{2.10}-\eqref{2.12} we see that $\C^2{\setminus}\{0\}$ coincides with the tautological line bundle over $\CPP^1$ (see e.g. \cite{Wells}),
\begin{equation}\label{2.13}
\C^2{\setminus}\{0\}=\CO(-1)^\times\ \stackrel{\C^*}{\longrightarrow}\ \CPP^1\ ,\quad \CO(-1)^\times =\CO(-1){\setminus}\{\psi_{-1}=0\}\ ,
\end{equation}
with the coordinates in fibres written in square brackets in \eqref{2.11} for $U_0$ and $U_1$. Here $\CO(-1)^\times$ denotes the bundle  $\CO(-1)$ without zero section $\psi_{-1}=0$. To add zero section, we need to go to the space $\C^2$ blowed-up at $\{0\}\in\C^2$, which we will consider in Section 7 in the context of zero-point energy and ground state. Note that the Hopf bundle \eqref{2.9} is a subbundle of unit vectors in the bundle $\CO(-1)$ given by the condition $|\psi_{-1}|\rho =1$. The fibres $S^1$ of the bundle \eqref{2.9} are parametrized by the coordinate $\vph$ in  \eqref{2.11}, \eqref{2.12}. When considering dynamics \eqref{2.8}, this coordinate is replaced by
\begin{equation}\label{2.14}
\vph (t)=\vph +\omega t\quad\Rightarrow\quad e^{\im\omega t}\ \ \mbox{in}\ \ z_\al (t)=e^{\im\omega t}z_\al\ ,
\end{equation}
and the initial value $\vph=\vph(0)$ can always be chosen equal to zero. The period of rotation of point \eqref{2.8}-\eqref{2.14} along the circle $S^1\hra S^3$ is equal to $T=2\pi/\omega$.

\section{Quantum line bundle}

\noindent{\bf QM as gauge theory}. In the geometric quantization approach it was shown that quantum mechanics is a gauge theory of special type on phase space of classical particles \cite{Sni, Wood, Hurt}. Quantization of a nonrelativistic system is a transition from phase manifold $X$ with a symplectic 2-form $\Omega_X$ ($=\dd\theta_X$ locally) to a principal U(1)-bundle $P(X, \sU(1)_{\sf v})$ over $X$ with a connection and curvature defined as
\begin{equation}\label{3.1}
\Av := -\frac{\im}{\hbar}\theta_X\und \Fv := -\frac{\im}{\hbar}\Omega_X\ .
\end{equation}
Both $\Av$ and $\Fv$ take values in the Lie algebra $\ru(1)_{\sf v}=\,$Lie$\sU(1)_{\sf v}$, where $\sU(1)_{\sf v}$ is the structure group of the above bundle. The abbreviations ``$\sf v$" and ``$\sf vac$" here mean ``vacuum" since the symplectic potential $\theta_X$ and the 2-form $\Omega_X$ (and, therefore, $\Av$ and $\Fv$) have {\it no sources}. Thus, we have a principal bundle $P(X, \sU(1)_{\sf v})$ over $X$ and a {\it background connection} $\Av$ on it.

The next step is to introduce a complex line bundle $\Lv$ associated with $P(X, \sU(1)_{\sf v})$ and to impose on sections of this bundle the constancy along integrable subbundle $\CT$ of the complexified tangent bundle $T^\C X$ of $X$ \cite{Sni, Wood, Hurt}. In the symplest case, this is the condition of independence either from momenta, or from coordinates, or holomorphicity of sections. Thus, quantum mechanics is a specific kind of Abelian gauge theory on phase space $X$ described by the set  $(\Lv\to X, \Av, \CT)$, where the connection $\Av= -\frac{\im}{\hbar}\theta_X$ is not dynamical. The curvature $\Fv$ of this connection defines the canonical commutation relations (CCR).

\smallskip

\noindent {\bf Quantum bundle}. In the case we are considering, $X=\C^2$, the bundle $\Lv$ is a complex 3-dimensional space with an obvious projection $\pi$ onto $\C^2$,
\begin{equation}\label{3.2}
\Lv=\C^2\times\C\ni (z_0,z_1,\Psi)\ , \quad \pi:\ \Lv\to\C^2\ ,
\end{equation}
with coordinates $z_\al$ on the base $\C^2$ and $\Psi$ on the fibre $\C$. The connection $\Av$ (gauge potential), given by the symplectic potential \eqref{2.5}, has components
\begin{equation}\label{3.3}
A_{z_\al}^{}=\sfrac12\,\delta^{\al\dal}\zb_\dal \und A_{\zb_\dal}^{}=-\sfrac12\,\delta^{\dal\al}z_\al\ .
\end{equation}
Accordingly, the covariant derivatives on the bundle \eqref{3.2} have the form
\begin{equation}\label{3.4}
\nabla_{z_\al}^{}=\dpar_{z_\al}^{}+A_{z_\al}^{}=\dpar_{z_\al}^{}+\sfrac12\delta^{\al\dal}\zb_\dal \und
\nabla_{\zb_\dal}^{}=\dpar_{\zb_\dal}^{}+A_{\zb_\dal}^{}=\dpar_{\zb_\dal}^{}-\sfrac12\delta^{\dal\al}z_\al \ ,
\end{equation}
with $\Av$ from \eqref{3.3}.

In the complex Bargmann-Fock-Segal representation \cite{Bar, Segal, Hall} the bundle $\Lv$ must be holomorphic. 
To pass from a complex bundle \eqref{3.2} with a fibrewise Hermitian metric $\bar\Psi\Psi$ to a holomorphic bundle, one must impose the holomorphicity condition on sections $\Psi$. To do this, we impose this condition on the ground state $\psi_0^{\sf v}$ using the Dolbeault operator 
$D_{\zb_\dal}^{}$, obtaining
\begin{equation}\label{3.5}
D_{\zb_\dal}^{}\psi_0^{\sf v}=\left(\dpar_{\zb_\dal}^{}+\sfrac12\delta^{\dal\al}z_\al \right)\psi_0^{\sf v}=0\quad\Rightarrow\quad
\psi_0^{\sf v}=|\psi_0^{\sf v}|e^{\im\theta}=\sfrac1\pi\, e^{-\frac12|z|^2}e^{\im\theta}\ ,
\end{equation}
where $|z|^2$ is defined in \eqref{2.2}.
The function $\psi_0^{\sf v}$ from \eqref{3.5} defines the ground state of the oscillator, which we will discuss in detail in Section 7. 
Accordingly, the holomorphic sections of the bundle $\Lv$ have the form
\begin{equation}\label{3.6}
\Psi=\psi (z)\psi_0^{\sf v}(z, \zb, \theta)\quad\Rightarrow\quad\bar\Psi\Psi=\frac1{\pi^2}\bar\psi\psi  e^{-|z|^2}\ ,
\end{equation}
where $\psi (z)$ is a holomorphic function of $z\in\C^2$.

\smallskip

\noindent {\bf Curvature $\Fv$ and CCR}. It is easy to see that the covariant derivatives \eqref{3.4} act on sections \eqref{3.6} as follows:
\begin{equation}\label{3.7}
\nabla_{z_\al}^{}(\psi\psi_0^{\sf v})=(\dpar_{z_\al}^{}\psi )\psi_0^{\sf v}\und 
\nabla_{\zb_\dal}^{}(\psi\psi_0^{\sf v})=-(\delta^{\dal\al} z_\al\psi)\psi_0^{\sf v}\ .
\end{equation}
Hence, we can introduce annihilation and creation operators by formulae
\begin{equation}\label{3.8}
a^\al :=\nabla_{z_\al}^{}\ ,\quad a_\beta^\+:=-\delta_{\beta\dot\beta}\nabla_{\zb_{\dot\beta}}^{}\quad\Rightarrow\quad
[a^\al, a_\beta^\+]=-F^{\al\dot\beta}\delta_{\dot\beta\beta}=\delta_\beta^\al\ .
\end{equation}
If we consider their action on $\psi(z)$ in \eqref{3.7}, transferring the function $\psi_0^{\sf v}$ to the integration measure as in \eqref{3.6}, then this will be the standard operators $a^\al =\dpar_{z_\al}^{}, a_\beta^\+={z_\beta}^{}$, and we will obtain the standard form for the CCR \eqref{3.8}. In fact, the commutator of covariant derivatives in \eqref{3.8} defines the curvature  $\Fv=F^{\al\dal}\dd z_\al{\wedge}\dd\zb_\dal$ of the bundle $\Lv$. Thus, ``quantum" coordinates and momenta, as well as creation and annihilation operators, are nothing more than covariant derivatives in the bundle $\Lv$. The non-commutativity of this ``quantum" operators follows from the non-zero curvature $\Fv$ of this bundle. We put the word ``quantum" in quotes because there is nothing misteriously quantum about the covariant derivatives on vector bundles used in differential geometry.

\smallskip

\noindent {\bf  Inner product.} Let us write out the inner product for functions of the form \eqref{3.6},
\begin{equation}\label{3.9}
\langle\Psi_1,\Psi_2\rangle =\langle\psi_1,\psi_2\rangle = \frac1{\pi^2}\int_{\C^2}\bar\psi_1(z)\psi_2(z)e^{-|z|^2}\dd^2z\ .
\end{equation}
In this complex representation the Hilbert space $\CH$ is introduced as the space of holomorphic functions on $\C^2$ square integrable with \eqref{3.9}, i.e. a function $\psi(z)$ belongs to $\CH$ if it has a finite norm $\|\psi\|=\langle\psi, \psi\rangle^{1/2}$. Let us emphasize once again that the whole quantumness is connected with the non-zero curvature $\Fv$ of the bundle \eqref{3.2} and the interaction of sections $\Psi$
 of the bundle $\Lv$ with the background gauge field $\Av$  through the covariant derivatives \eqref{3.4} and \eqref{3.7}.

\section{Quantum oscillator}

\noindent {\bf Dynamics.} After describing the Hilbert space $\CH$, it is necessary to determine the dynamics of the change of the wave functions $\Psi$ over time $t$. The Hamiltonian \eqref{2.2} of the classical oscillator is mapped into a covariant Laplacian
\begin{equation}\label{4.1}
\hat H =-\hbar\omega\Delta_2=-\sfrac12\,\hbar\omega\,\delta_{\al\dal}\bigl(\nabla_{z_\al}^{}\nabla_{\zb_{\dal}}^{}+
\nabla_{\zb_{\dal}}^{}\nabla_{z_\al}^{}\bigr)=-\hbar\omega\,\delta_{\dal\al}\nabla_{\zb_{\dal}}^{}\nabla_{z_\al}^{}+\hbar\omega\ ,
\end{equation}
acting on sections $\Psi$ of the bundle \eqref{3.2}. Note that the term $\hbar\omega$ in \eqref{4.1}  arrose from the commutator of covariant derivatives. On holomorphic sections \eqref{3.6} we have 
\begin{equation}\label{4.2}
\hat H\Psi =(\sfrac\hbar\im V_H\psi)\psi_0^{\sf v} + \hbar\omega\psi\psi_0^{\sf v}=\hbar\omega(z_\al\dpar_{z_\al}^{}\psi + \psi)\psi_0^{\sf v}\ ,
\end{equation}
where $V_H$ is the generator of the group U(1) introduced in \eqref{2.6}.

For an arbitrary holomorphic function 
\begin{equation}\label{4.3}
\Psi(z)=\psi(z)\psi_0^{\sf v}=\psi_0^{\sf v}\sum_{n=0}^{\infty} c_n\psi_n(z)\for\psi_n(z)=\sum_{m=0}^{n}b_{nm}\frac{z_0^{n-m}z_1^m}{\sqrt{(n-m)!m!}}\ ,
\end{equation}
the evolution is given by the Schr\"odinger equation
\begin{equation}\label{4.4}
\im\,\hbar\,\dpar_t\Psi(t)=\hat H\Psi (t)=\left(-\hbar\omega\delta_{\dal\al}\nabla_{\zb_{\dal}}^{}\nabla_{z_\al}^{}+\hbar\omega\right)\Psi(t)
\end{equation}
with solution
\begin{equation}\label{4.5}
\Psi(t)=e^{-\frac\im\hbar t\hat H}\Psi=\Bigl(\sum_{n=0}^{\infty} c_n\psi_n e^{-\im\omega nt}\Bigr)(\psi_0^{\sf v}e^{-\im\omega t})\ .
\end{equation}
Note that the zero-point energy $E_0^{\sf v}=\hbar\omega$ is related to the ground state $\psi_0^{\sf v}(t)=\exp(-\im\omega t)\psi_0^{\sf v}$ rotations, which we will discuss in detail in Section 7. Recall that the term $E_0^{\sf v}$ arises in \eqref{4.1} due to the non-commutativity of the covariant derivatives \eqref{3.4} with constant curvature $\Fv$ in \eqref{3.8}. The behavior of particles in such an Abelian field is identical to the behavior of electrically charged particles in a constant magnetic field (Landau levels).

\smallskip

\noindent {\bf  Charge density.} It is easy to show that from \eqref{4.4} follows the continuity equation,
\begin{equation}\label{4.6}
\dpar_t\rho_{\sf v} + \nabla_{z_\al}^{}j_\al + \nabla_{\zb_{\dal}}^{}j_\dal =0\ ,
\end{equation}
where 
\begin{equation}\label{4.7}
\rho_{\sf v}=\bar\Psi\Psi=\frac{1}{\pi^2}\bar\psi\psi e^{-|z|^2}\ ,\quad 
j_\al =\im\omega\delta_{\al\dal}\bigl(\bar\Psi\nabla_{\zb_{\dal}}^{}\Psi - (\nabla_{\zb_{\dal}}^{}\bar\Psi)\Psi\bigr)=-\bar j_\dal\ .
\end{equation}
According to the gauge theory view discussed in this paper, $\rho_{\sf v}$ is the density of quantum charge $q_{\sf v}=1$ of section $\Psi$ of the bundle $\Lv$. In the standard approach, the function $\rho_{\sf v}$ is associated with probabilities, although not as directly as in the coordinate representation. These views on $\rho_{\sf v}$ are not contradictory.

\smallskip

\noindent {\bf Grading of Hilbert space $\CH$}. On holomorphic functions $\psi(z)$, the operator $\hat H$ from \eqref{4.1} without $E_0^{\sf v}$ reduces to the operator 
\begin{equation}\label{4.8}
\hat N = z_\al\dpar_{z_\al}^{}\ ,\quad \hat N\psi_n=n\psi_n\ ,\quad E_n=\hbar\omega n\ .
\end{equation}
This operator defines a {\it grading} of the Hilbert space $\CH$ (see e.g. \cite{Thomas}),
\begin{equation}\label{4.9}
\psi(z)=\sum_{n=0}^{\infty} c_n\psi_n\ \in\ \CH=\mathop\oplus_{n=0}^{\infty}\C^{n+1}\ ,
\end{equation}
where homogeneous polynomials $\psi_n$ of degree $n$ from \eqref{4.3} parametrize irreducible representations $\C^{n+1}$ of the group SU(2) of spin $j=\sfrac12n$. Note that here we consider the Hilbert space $\CH$ without the multiplier $\psi_0^{\sf v}$ from \eqref{4.3} transferring it to the definition of the inner product \eqref{3.9}, as it is accepted in the Bargmann-Fock-Segal approach \cite{Bar, Segal, Hall}. The time dependence is introduced by operator $\exp(-\sfrac\im\hbar\,t\,\widehat N)$ and is written out in formula \eqref{4.5}. The orthomormal basis in the Hilbert space $\CH$ with respect to the inner product \eqref{3.9} is formed by the functions
\begin{equation}\label{4.10}
\psi_{nm}(z)=\frac{z_0^{n-m}z_1^m}{\sqrt{(n-m)!m!}}\ ,\quad m=0,...,n,\ n=0,1,... ,
\end{equation}
from which it is clear that $\psi_0=b_{00}\psi_{00}=1$. For a fixed value of $n\in \Nbb$, all $n+1$ functions \eqref{4.10} have the same energy value $E_n$ (degeneracy) and can therefore be combined into an eigenfunction $\psi_n(z)$ from \eqref{4.3}. It is these functions that will be the subject of our further analysis.

\section{Mathematical nature of $\psi_n$}

\noindent{\bf Wave function $\psi_n(z)$.}  In quantum mechanics it is postulated that the physical state of a system is completely described by the wave function \eqref{4.5} and everything that can be known about a quantum particle is contained in $\Psi$. We postpone the discussion of the ground state $\psi_0^{\sf v}$ from \eqref{3.5}-\eqref{3.7} and \eqref{4.2}-\eqref{4.5} until Section 7 and consider here the eigenfunctions of the grading operator \eqref{4.8},
\begin{equation}\label{5.1}
\psi_n(z)=\sum_{m=0}^nb_{nm}\psi_{nm}(z)\ .
\end{equation}
Here $\psi_0=1$ and we need to clarify what information is contained in $\psi_n$ with $n\in\Nbb$.

\smallskip

\noindent {\bf Weight $n$.} The function \eqref{5.1} is a section of a special type of trivial bundle \eqref{3.2}. This function $\psi_n: \C^2\to\C\subset\Lv$ is a homogeneous polynomial of degree $n$, that is, when $z$ is replaced by $\lambda z$, it changes as $\psi_n(\lambda z)=\lambda^n\psi_n(z)$, where $\lambda\in\C^*=\sGL(1,\C)$. This property means that $\psi_n$ belongs to the space $\C_{(n)}$ of the one-dimensional holomorphic representation of the group $\C^*$ of weight $n$
\begin{equation}\label{5.2}
\nu :\quad\C^*\to\sGL(1,\C)\ ,\quad \nu(\lambda)\cdot\psi_n=\lambda^n\psi_n\ ,\quad\lambda\in\C^*\ .
\end{equation}
Note that the property \eqref{5.2} is preserved in the case where $\psi_n$ is not a section, but a coordinate on a fibre $\C_{(n)}$ in the space
\begin{equation}\label{5.3}
Y_n:=(\C^2{\setminus}\{0\})\times \C_{(n)}\ \ni\ (z_0, z_1, \psi_n)
\end{equation}
embedded in the bundle $\Lv$. Here $z_\al$ have weight $+1$.

\smallskip

\noindent {\bf Almost quantum}. We will call the consideration of the motion of a point in the total spaces of the bundles $\Lv$ and $Y_n$ over $\C^2$ almost quantum mechanics. The dynamics of points in phase space $\C^2$ is a problem of classical mechanics. Considering the dynamics of points in a space $\Lv$ that has an additional complex coordinate compared to $\C^2$ is not classical mechanics, but also not quantum mechanics, where points are replaced by sections (wave functions). The graph of a section is a phase space $\C^2$ embedded in the space $\Lv$, all points of which move within the total space $\Lv$. We will further consider the almost quantum case, and later we will move to the quantum case, emphasizing what kind of additional information  is added when moving from points to sections.

\smallskip

\noindent {\bf Representations of $\C^*$}. The transition from the coordinates on the bundle to sections and back to the coordinates on the bundle can be understood as follows. The coordinates $(z_0, z_1)$ admit the action of the group $\C^*\times\C^*$ with weight $+1$, where the first group acts on $z_0$, and the second group acts on $z_1$. Let us introduce additional coordinates $(y_0, y_1)$ of weight $+1$ relative to the action of the same group $\C^*\times\C^*$. Then the monomial
\begin{equation}\label{5.4}
\psi_{nm}(y)=\frac{y_0^{n-m}y_1^m}{\sqrt{(n-m)!m!}}\ ,\quad m=0,...,n,\ n=0,1,... ,
\end{equation}
has weight $(n-m,m)$ with respect to the action of the group $\C^*\times\C^*\ni (\lambda_0, \lambda_1)$. We define a diagonal subgroup $\C^*=\diag (\C^*\times\C^*)$ with $\lambda_0=\lambda_1=\lambda$. Then the function
\begin{equation}\label{5.5}
\psi_n(y)=\sum_{m=0}^nb_{nm}\psi_{nm}(y)\quad\Rightarrow\quad\psi_n(\lambda y_0, \lambda y_1)=\lambda^n\psi_n(y_0,y_1)
\end{equation}
has weight $n$ with respect to the diagonal group $\C^*$, i.e. $\psi_n\in\C_{(n)}$ according to \eqref{5.2}.

The coordinates $\psi_{nm}$  and $\psi_n$ on the fibres of the bundles $Y_{nm}\ni (z_0, z_1, \psi_{nm})$ and $Y_{n}\ni (z_0, z_1, \psi_{n})$ can be understood as functions \eqref{5.4} and \eqref{5.5} of the additional coordinates $y_\al$. In this case, on the space $Y_n$ the action \eqref{5.5} of the group $\C^*$ is defined together with $z_\al\mapsto\lambda z_\al$. In addition to this, on the space $Y_{nm}$ there is defined the action of the group $\C^*$ of the form $y_1\mapsto\lambda_1 y_1$, $\psi_{nm}\mapsto\lambda_1^m\psi_{nm}$. These actions of the two groups $\C^*$ define the spaces $Y_n$ and $Y_{nm}$ as trivial equivariant bundles over the phase space $\C^2$. The transition from coordinates on fibres of these bundles to cross sections is accomplished by replacing $y_\al$ with $z_\al$.

\smallskip

\noindent {\bf $\CO(n)$ in GIT approach.} On the space $Y_n$ from \eqref{5.3} the action of the group $\C^*$ is defined:
\begin{equation}\label{5.6}
\lambda\cdot (z_0, z_1, \psi_n)=(\lambda z_0, \lambda z_1, \lambda^n\psi_n)\for \lambda\in\C^*\ .
\end{equation}
It is well known that the bundle \eqref{5.3} is a pull-back of the bundle $\CO(n)\to\CPP^1$ from the Riemann sphere $\CPP^1\hra\C^2{\setminus}\{0\}$ to $\C^2{\setminus}\{0\}$, which we will now consider. Note that time $t$ is an external parameter and in \eqref{5.1}-\eqref{5.6} we should write $\psi_n\exp(-\im\omega nt)$ but for now we are considering $t=0$.

The Geometric Invariant Theory (GIT) is an approach to constructing quotient space $Y/G$ where an algebraic group $G$ acts on an algebraic variety $Y$ \cite{Mumford, Thomas}. In the case under consideration, the variety $Y_n$ is defined in \eqref{5.3}, $G=\C^*$ and the action of $\C^*$ on $Y_n$ is defined in \eqref{5.6}. The bundle $\CO(n)$ is defined as an equivalence class $Y_n/\C^*$ under the action \eqref{5.6} of the group $\C^*$,
\begin{equation}\label{5.7}
(\lambda z_0, \lambda z_1, \lambda^n\psi_n)\sim (z_0, z_1, \psi_n)\ .
\end{equation}
In fact, the equivalence of the points in \eqref{5.7} means that $\C^*$ is a ``gauge group" and to specify the coordinates of the complex two-dimensional  manifold $\CO(n):=Y_n/\C^*$ one has to ``fix a gauge".

Gauge is fixed differently on patches $U_0$ and $U_1$ from \eqref{2.10},
\begin{equation}\label{5.8}
\begin{split}
U_0:\ z_0\ne 0,\ \lambda =z_0^{-1}\quad&\Rightarrow\quad (\lambda z_0, \lambda z_1, \lambda^n\psi_n)=(1, z, \psi_nz_0^{-n})\ ,
\\[1pt]
U_1:\ z_1\ne 0,\ \lambda =z_1^{-1}\quad&\Rightarrow\quad (\lambda z_0, \lambda z_1, \lambda^n\psi_n)=(w, 1, \psi_nz_1^{-n})\ .
\end{split}
\end{equation}
By choosing $\lambda$ in \eqref{5.8} we select a {\it representative} of the equivalence class that fixes the admissible transformations from group $\C^*$ to the identity, completely eliminating excess degrees of freedom. On the intersection $U_0\cap U_1$ of patches \eqref{5.8} we have gluing by transition function $z^n$ of the bundle $\CO(n)$:
\begin{equation}\label{5.9}
\psi_nz_0^{-n}=z^n(\psi_nz_1^{-n})\ ,\quad z=\frac{z_1}{z_0}\ .
\end{equation}
We have obtained a covering of the manifold $\CO(n)$ by two charts with two complex coordinates on $U_0\times\C$ and two on $U_1\times\C$. The Chern class of this bundle is given by the degree $n$ of the transition function, $c_1(\CO(n))=n$. We introduce a holomorphic line bundle $\CO(-n)$ dual to $\CO(n)$ with fibre coordinates and transition function of the form
\begin{equation}\label{5.10}
\psi_{-n}z_0^{n}=z^{-n}(\psi_{-n}z_1^{n})\ ,
\end{equation}
setting $\psi_{-n}:=\psi_n^{-1}$. The Chern class of this bundle is $c_1(\CO(n))=-n$.

\smallskip

\noindent {\bf Lens spaces.} In formulae \eqref{5.8}-\eqref{5.10} we introduced $\CO(n)$ and $\CO(-n)$ as holomorphic line bundles over $\CPP^1$, and now we redefine them as Hermitian bundles, setting
\begin{equation}\label{5.11}
\begin{split}
U_0:\ z_0\ne 0,\ \lambda =\Bigl(\frac{\zb_{\dot 0}}{z_0}\Bigr)^{1/2}\rho^{-1}\quad&\Rightarrow\quad \Bigl(\frac{1}{\sqrt{1+ z\zb}}, \frac{z}{\sqrt{1+ z\zb}}, \psi_n\rho^{-n}e^{-\im\vph_0n}\Bigr)\ ,
\\[3pt]
U_1:\ z_1\ne 0,\ \lambda =\Bigl(\frac{\zb_{\dot 1}}{z_1}\Bigr)^{1/2}\rho^{-1}\quad&\Rightarrow\quad \Bigl(\frac{w}{\sqrt{1+ w\bar w}}, \frac{1}{\sqrt{1+ w\bar w}}, \psi_n\rho^{-n}e^{-\im\vph_1n}\Bigr)\ ,
\end{split}
\end{equation}
where $\vph_0, \vph_1$ and $\rho$ are written out in \eqref{2.1}, \eqref{2.2} and \eqref{2.12}. For the bundle $\CO(-n)$, one should take the fibre coordinates that are inverse to the fibre coordinates in \eqref{5.11}.

Let us now consider subbundles of vectors of fixed length (for example, unit length) in the fibres $\C$ of bundles $\CO(n)$ and $\CO(-n)$,
\begin{equation}\label{5.12}
L(n,1):=S(\CO(n)):\quad  \frac{\psi_n}{|\psi_n|}e^{-\im\vph n}e^{\im\chi n/2}= e^{\im\chi n}\Bigl(\frac{\psi_n}{|\psi_n|}e^{-\im\vph n}e^{-\im\chi n/2}\Bigr)\ ,
\end{equation}
\begin{equation}\label{5.13}
L(-n,1):=S(\CO(-n)):\quad  \frac{|\psi_n|}{\psi_n}e^{\im\vph n}e^{-\im\chi n/2}= e^{-\im\chi n}\Bigl(\frac{|\psi_n|}{\psi_n}e^{\im\vph n}e^{\im\chi n/2}\Bigr)\ ,
\end{equation}
where the angles $\vph$ and $\chi$ are given in \eqref{2.12}, $n\geq 1$. On the left and right in formulae \eqref{5.12}, \eqref{5.13} we have written out the coordinates on the fibres $S^1\subset\C_{(n)}$ of these bundles over $U_0$ and $U_1$ in $\CPP^1$, with transition function $(z/\zb)^{n/2}$ for the Chern class $c_1=n$ and  $(\zb/z)^{n/2}$ for the Chern class $c_1=-n$. The case $n=0$ with $\psi_0=1$ is special -- here the coordinate $\psi_0^{\sf v}=|\psi_0^{\sf v}|\exp(\im\theta)$ from \eqref{4.5}    is defined on the fibre $\C_{(0)}$ of the trivial bundle, 
\begin{equation}\label{5.14}
\CO(0)=\CPP^1\times\C_{(0)}\supset L(0,1):=\CPP^1\times S^1\ ,\quad \exp(\im\theta)\in S^1\ .
\end{equation} 
We will discuss this case in Section 7.

\smallskip

\noindent {\bf Dynamics in lens spaces.}   The bundles $L(\pm n, 1)$  are principal U(1)-bundles over $\CPP^1$ embedded in $\CO(\pm n)$ as 3-dimensional boundaries of their associated disk bundles. In the next section we will show that as manifolds they are quotient spaces of the Hopf sphere \eqref{2.9} by the action of the cyclic group $\Z_n$ with $n\geq 1$,
\begin{equation}\label{5.15}
L(\pm n, 1)\cong S^3/\Z_n\ ,\quad \pi_1(S^3/\Z_n)=\Z_n\ ,
\end{equation} 
and cases $n$ and $-n$ differ only in orientation. This difference can be seen in formulae \eqref{5.12} and \eqref{5.13} as a change in the signs of the angles to the opposite ones. To describe a particle moving in these bundles, $\psi_n$ in \eqref{5.12} and \eqref{5.13} should be replaced by $\psi_n(t)=\psi_n\exp(-\im\omega nt)$. Then, for $n\geq 1$, we will see a particle moving in a circle in the fibre $S^1$ as
\begin{equation}\label{5.16}
\exp(\mp\im(\vph +\omega t)n)\in S^1\ \hra\  S^3/\Z_n\stackrel{S^1}{\longrightarrow}\CPP^1\ ,
\end{equation} 
and for $L(n,1)$ and $L(-n,1)$ it moves in opposite directions with a period
\begin{equation}\label{5.17}
T_n=\frac{2\pi}{\omega n}\ .
\end{equation} 
Recall that the energy of this motion is equal to $E_n=\hbar\omega n$, and the phase $\vph$ and $[z_0{:}z_1]\in \CPP^1$ define the initial data of this motion. Energy $E_0^{\sf v}=\hbar\omega$ is the energy of motion of a particle in fibre $\C_{(0)}$ of the bundle \eqref{5.14}, which we will consider in Section 7. Let us recall that we are discussing the algebraic-geometric description of bundles and the motion of a point in them, that is, $\psi_n$ is a coordinate, not a function. Therefore, there are no probabilities yet, it is an almost quantum case. For the same reason, there is no factorization by global phase; the motion in question is real.

\smallskip

\noindent {\bf Group $\Z_n$.}  The cyclic group $\Z_n=\Z/n\Z$ of order $n$ in \eqref{5.16} is generated by an element $\zeta\in\C^*$ with $\zeta^n=1$, i.e. $\zeta$ is $n$-th root of unity,
\begin{equation}\label{5.18}
\Z_n=\bigl\{\zeta^\ell =\exp(\im\,\sfrac{2\pi\ell}n),\ \ell =0,...,n-1\mid\zeta=\exp(\sfrac{2\pi\im}n)\bigr\}\ .
\end{equation} 
This group $\Z_n\subset \sU(1)$ acts by discrete rotations on the circle $S^1$ in the spaces $\C_{(n)}$ from \eqref{5.3} and in the fibres of the bundles \eqref{5.11}-\eqref{5.16}, 
\begin{equation}\label{5.19}
\vph\mapsto\vph + \sfrac{2\pi}n\quad \Rightarrow\quad \vph_n=\vph n\quad \Rightarrow\quad \vph_n\mapsto\vph _n+ {2\pi}\ ,
\end{equation} 
and we see that the particle makes one revolution around the circle $S^1$ in time \eqref{5.17}.

\smallskip

\noindent {\bf Sections of $\CO(n)$}. We have shown that local coordinates in the fibres of the bundles \eqref{5.8} and \eqref{5.14} carry information about the Chern class of the bundles $\CO(\pm n)$ and about the fundamental group $\Z_n$ of the lens spaces \eqref{5.12}, \eqref{5.13} and  \eqref{5.15}.  The trajectory of point in \eqref{5.16} depends on the point $z\in\CPP^1$ on the sphere $\CPP^1$, which specifies the initial data for the motion. This is not a quantum motion yet.

The quantum case arises when, instead of a point in a fibre above a specific point $z\in\CPP^1\subset\C^2{\setminus}\{0\}$, we take a {\it section} of a bundle, that is, we consider points in fibres above all points of the base of the bundle simultaneously. This means that we are considering mappings
\begin{equation}\label{5.20}
\psi_n:\ \CPP^1\to\CO(n)\und\psi_n:\ \C^2{\setminus}\{0\}\to Y_n\ ,
\end{equation} 
where the first mapping follows from the second and vice versa. In terms of the representation \eqref{5.2}-\eqref{5.5} of the group $\C^*\times\C^*$, to go from fibre coordinate $\psi_n$ to section $\psi_n(z)$, one should simply replace $y_\al$ with $z_\al$ in \eqref{5.5}.

The polynomial $\psi_n(z_0, z_1)$ on $\C^2$ is given in \eqref{4.10} and \eqref{5.1}, and on $\CPP^1=U_0\cup U_1$ it is given through local sections
\begin{equation}\label{5.21}
U_0:\ \psi_n(z)=\psi(z_\al)z_0^{-n}=\sum_{m=0}^n b_{nm}z^m\ ,
\quad U_1:\ \psi_n(w)=\psi(z_\al)z_1^{-n}=\sum_{m=0}^n b_{nm}w^{n-m}
\end{equation}
glued at the intersection $U_0\cap U_1$ according to formula \eqref{5.9}. These functions, as well as functions $\psi_n$ in \eqref{5.1}, define global holomorphic sections of the bundle $\CO(n)$, the coefficients $\{b_{nm}, m=0,...,n\}$ of which parametrize the irreducible representations $\C^{n+1}$ of the group SU(2), $n\in\Nbb$. The dual bundle $\CO(-n)$ has no global holomorphic sections, and we define its meromorphic section by the function $\psi_{-n}=\psi_n^{-1}$ inverse to $\psi_n$. The choice of meromorphic sections in $\CO(-n)$ is not unique, but the difference between the number of zeros and poles of a section must be equal to the number $c_1(\CO(-n))=-n$. We take a section $\psi_{-n}=\psi_n^{-1}$ that has only poles.

What happens when instead of the point \eqref{5.6} in the bundle $\CO(n)$ we take a section?  The coordinate $z$ on $\CPP^1\subset S^3/\Z_n$ was the initial value for the trajectory of point \eqref{5.16} passing through it. For a section, which is an embedding of $\CPP^1$ into the total space $\CO(n)$, all points of $\CPP^1$ move synchronously along the circle $S^1$ in \eqref{5.16}. Moreover, on the fibre of the bundle $\CO(n)$ over each point $z$, the metric $\bar\psi_n\psi_n\exp(-|z|^2)$ is defined, i.e. the points $z$ have different ``weights". The Born postulate assigns a probabilistic meaning to this weight, although in the holomorphic representation it is more complicated than in the coordinate representation.

\smallskip

\noindent {\bf Majorana stars.} Note that the function $|\psi_n|^2$ on sections has singular points which we will now describe. The polynomial \eqref{5.21} has $n$ zeros at points $P_i$ on the sphere $\CPP^1\hra\CO(n)$. Some points $P_i$ with different $i$ may coincide and then they are taken into account with multiplicity. We introduce a formal sum $D_n$ of points $P_i$ on $\CPP^1$,
\begin{equation}\label{5.22}
D_n=\sum_{i=1}^n P_i\ ,
\end{equation}
called a {\it divisor}. The degree of a divisor is equal  to the sum of its coefficients ($+1$ for zero and $-1$ for a pole), so for \eqref{5.22} we have deg$(D_n)=n$. For the bundle $\CO(-n)$ we define the divisor as $D_{-n}=-D_n$ with deg$(D_{-n})=-n$. Note that divisors completely determine the bundles $\CO(\pm n)$. 

It is interesting that in physics the points $P_i$ (zeros of the polynomials $\psi_n$) are called Majorana stars, and the divisor \eqref{5.22} is called the Majorana constellation. These concepts, characterizing the properties of system with spin, are actively used in quantum optics and other applied areas of physics.

\smallskip

\noindent {\bf Graph of a section.} The functions \eqref{5.21} define an embedding of the base $\CPP^1$ into the total space of the bundle $\CO(n)$ (and into $S^3/\Z_n$ through \eqref{5.12}), and we denote this embedded sphere by $\CPP^1_{(n)}\hra\CO(n)$. Locally it is defined by the graphs of functions
\begin{equation}\label{5.23}
\CPP^1_{(n)}:\ (z, \psi_n(z))\ \mbox{on}\ U_0\und (w, \psi_n(w))\ \mbox{on}\ U_1\ .
\end{equation}
This embedded sphere coincides with the base $\CPP^1$ of the bundle $\CO(n)$ at $\psi_n=0$ (zero section), i.e. at $n$ points $P_i\in\CPP^1$ (taking into account the multiplicity).  Thus, the Chern class $n$ also determines the self-intersection index of the general section \eqref{5.23} of the bundle $\CO(n)$. The dynamics of the section \eqref{5.23} is given by multiplying the function $\psi_n$ by $\exp(-\im\omega nt)$, i.e. all points of the sphere $\CPP^1_{(n)}$ rotate.

For the bundle $\CO(-n)$, the sections $\CPP^1_{(-n)}$ have the same singular points $P_i\in\CPP^1$, but the meromorphic section $\psi_{-n}=\psi_n^{-1}$ in them goes to infinity. Over the punctured base $\CPP^1{\setminus}\{P_1,...,P_n\}$, the bundles $\CO(n)$ and $\CO(-n)$ are trivial, and the topology of these bundles describes the behavior of fibres as they pass around the punctures. It is the bypass of these $n$ points $P_i\in\CPP^1$ that lead to phases $\exp(-\sfrac\im 2\chi n)$ and $\exp(\sfrac\im 2\chi n)$  in  \eqref{5.12} and  \eqref{5.13}. To sum up, the eigenfunction $\psi_n\exp(-\im\omega nt)$ of the operator $\hat N$ defines a point moving along a circle $S^1$ in the lens space $L(n,1)\subset\CO(n)\subset\Lv$. By replacing the point with the section $\CPP^1_{(n)}\subset L(n,1)\subset\CO(n)\subset \Lv$, we obtain the rotation of the entire sphere $\CPP^1_{(n)}$ of the initial data as a single whole.

\section{$\Z_n$-invariant dynamics in phase space}

\noindent {\bf Physical gauge}. Recall that the manifold $\CO(n)$ was introduced in \eqref{5.6}-\eqref{5.9} as a coset space by the action of the group $\C^*$,
\begin{equation}\label{6.1}
Y_n\ \stackrel{\C^*}{\longrightarrow}\  Y_n/\C^* =\CO(n)\ .
\end{equation}
Note that $\CO(0)=\CPP^1\times\C_{(0)}$ is a trivial bundle, and we defer the consideration of the case $n=0$ until Section 7. The manifold $\CO(n)$ with $n\geq 1$ was defined by the choice of gauge \eqref{5.8} or \eqref{5.11}. In this gauge we described the particle as a point moving along a circle $S^1\subset L(n,1)$ with period $T_n$ from \eqref{5.17}. In other words, in the almost quantum case, the particle moves in the manifold $\CO(n)$, which is a {\it representative} of the equivalence class \eqref{5.7}. But this is not the only representative of this equivalence class, and we will now move on to discuss it.

Note that if we set $\psi_n=0$ in \eqref{5.7}, then the equivalence relation defines the manifold $\CPP^1$,
\begin{equation}\label{6.2}
\psi_n=0:\ (\lambda z_0, \lambda z_1, 0)\sim(z_0, z_1, 0)\ \Rightarrow\ (1,z)\ \mbox{on}\ U_0\und (w,1)\ \mbox{on}\ U_1\ .
\end{equation}
This is the Riemann sphere embedded into $\CO(n)$ as a zero section. Next we will consider the manifold $\CO(n)$ without zero section, setting
\begin{equation}\label{6.3}
\CO(n)^\times :=\CO(n){\setminus}\{\psi_n=0\}\ .
\end{equation}
If $\psi_n\ne 0$, then in the space $Y_n$ from \eqref{5.3}, \eqref{5.6}-\eqref{5.8} we can choose another {\it representative} of the equivalence class \eqref{5.7} by setting
\begin{equation}\label{6.4}
\lambda=\psi_n^{-\frac1n}=\psi_{-n}^{\frac1n}\quad\Rightarrow\quad
 (\lambda z_0, \lambda z_1, \lambda^n\psi_n)=(\psi_{-n}^{\frac1n}z_0, \psi_{-n}^{\frac1n}z_1, 1)\ .
\end{equation}
By acting on \eqref{6.4} from the left with an element $\zeta\in\C^*$, we obtain
\begin{equation}\label{6.5}
 (\psi_{-n}^{\frac1n}\zeta z_0, \psi_{-n}^{\frac1n}\zeta z_1, \zeta^n)\sim
 (\psi_{-n}^{\frac1n}z_0, \psi_{-n}^{\frac1n}z_1, 1)\ ,
\end{equation}
which is satisfied if
\begin{equation}\label{6.6}
(\zeta z_0, \zeta z_1)\sim (z_0,z_1)\for \zeta^n=1\ \Rightarrow\ \zeta\in\Z_n\ .
\end{equation}
Thus, we have chosen another representative \eqref{6.6} of the equivalence class \eqref{5.7} and reduced the gauge group $\C^*$ to a finite subgroup $\Z_n\subset U(1)\subset\C^*$. The equivalence class \eqref{6.6} is a coset space $(\C^2{\setminus}\{0\})/\Z_n$ called an orbifold or conical space. Thus, we have a holomorphic isomorphism (biholomorphism),
\begin{equation}\label{6.7}
\CO(n)^\times\cong (\C^2{\setminus}\{0\})/\Z_n\ ,
\end{equation}
described by diagram \eqref{1.3} from the Introduction. We will consider the case $n=0$ in Section 7.

\smallskip

\noindent {\bf Local coordinates.} Note that $(\C^2{\setminus}\{0\})/\Z_n$ is a conical space with conical angle $2\pi/n$, generalizing the cone \eqref{2.3} corresponding to $n=1$. For any $n\in\Nbb$, the spaces \eqref{6.7} are flat, since the curvature is concentrated at the point $\{0\}\in\C^2$, which is fixed under the action of the group $\Z_n$. Furthermore, the singularity at this point is eliminated by blowing up the space $\C^2$ at origin, which we will discuss in Section 7.  The finite gauge group $\Z_n$ remains in  \eqref{6.7} because the equation $\lambda^n=\psi_{-n}$ has $n$ solutions (multivaluedness) and the choice of the branch of the $n$-th degree root is changed by multiplying by $\zeta\in\Z_n$.

As we have already discussed in \eqref{5.19}, the group $\Z_n$ acts by shifting  the angle $\vph$ by $2\pi/n$. To see this action explicitly, we introduce spherical coordinates  on the conical space  $(\C^2{\setminus}\{0\})/\Z_n$ similar to formula \eqref{2.11},
\begin{equation}\label{6.8}
\begin{split}
U_0:\ z_0\ne 0\quad\Rightarrow\quad   (\psi_{-n}^{\frac1n}z_0, \psi_{-n}^{\frac1n}z_1)=&[\psi_{-n}^{\frac1n}\rho e^{\im\vph}e^{-\im\chi/2}]\frac{(1,z)}{\sqrt{1+z\zb}}  \ ,
\\[3pt]
U_1:\ z_1\ne 0\quad\Rightarrow\quad (\psi_{-n}^{\frac1n}z_0, \psi_{-n}^{\frac1n}z_1)=&[\psi_{-n}^{\frac1n}\rho e^{\im\vph}e^{\im\chi/2}]\frac{(w,1)}{\sqrt{1+w\bar w}}     \ .
\end{split}
\end{equation}
As we see, $\psi_{-1}$ from  \eqref{2.11} is replaced  by $\psi_{-n}^{\frac1n}$, $n\in\Nbb$. The transition function for the coordinates in square brackets here is still $z^{-1}$ in the holomorphic basis and $\exp(-\im\chi)$ in the unitary basis. The difference from the bundle $\CO(-1)$ from \eqref{2.13} is that, due to the equivalence $\vph +2\pi/n\sim\vph$ given by the group $\Z_n$, the particle moves along the circle $\exp(\im\vph )$ with the period $T_n=2\pi/\omega_n$ from 
\eqref{5.17}, that is, $n$ times faster than in case \eqref{2.14}. The equivalence of the particle motion in two isomorphic spaces \eqref{6.7} leads to the fact that for $\Z_n$-invariant motion of a particle in $(\C^2{\setminus}\{0\})$ its energy is equal to $E_n=\hbar\omega n$.

\smallskip

\noindent {\bf Conical spaces.} Note that the orbifold \eqref{6.6}-\eqref{6.8} is a bundle $\CO(-1)^\times$ from \eqref{2.13} factorized by the group $\Z_n$, i.e. $\CO(-1)^\times/\Z_n$. Formally, this space can be represented as a fibre bundle over $\CPP^1$ with fibre $\C^*/\Z_n$. The transition to this space can be described by the chain
\begin{equation}\label{6.9}
\bigl\{\CO(n), \psi_nz_0^{-n}\, \&\, \psi_nz_1^{-n}\bigr\}\to
\bigl\{\CO(-n), \psi_{-n}z_0^{n}\, \&\, \psi_{-n}z_1^{n}\bigr\}\to
\bigl\{\sqrt[n]{\CO(-n)}, \psi_{-n}^{\frac1n}z_0\, \&\, \psi_{-n}^{\frac1n}z_1\bigr\} ,
\end{equation}
where $\CO(-1)/\Z_n$ can be formally represented as the $n$-th root of the bundle $\CO(-n)$. The operation of taking the $n$-th root of a bundle is not defined in the category of manifolds, but is well defined in the more general category of stacks. For the purposes of this paper, there is no need to describe the orbifold $(\C^2{\setminus}\{0\})/\Z_n$ as a bundle over $\CPP^1$, it is enough to describe it as a manifold, so there is no need for stacks. To describe the orbifold 
$(\C^2{\setminus}\{0\})/\Z_n$ as a manifold, it is sufficient to fix the choice of a {\it representative} of the equivalence class \eqref{6.6} by setting
\begin{equation}\label{6.10}
0\leq\vph +\omega t <\frac{2\pi}n\quad\Rightarrow\quad 0\leq n(\vph +\omega t) <{2\pi}
\end{equation}
and identifying the boundary points. This exactly corresponds to phase $\vph_n=\vph n$ in the bundles $\CO(\pm n)$ from \eqref{5.12},\eqref{5.13}. It is this second option that is used in physics (see e.g. \cite{SS, BV, Bordag}). Note that orbifolds and conical spaces are often considered in many areas of physics.

\smallskip

\noindent {\bf Equivalent descriptions.} Fixing $\rho|\psi_{-n}^{\frac1n}|$ in \eqref{6.8} to a positive constant, we obtain a lens space \eqref{5.12} embedded in the Hopf sphere \eqref{2.9}-\eqref{2.12} as a submanifold distinguished by condition \eqref{6.6} and \eqref{6.10} with projection
\begin{equation}\label{6.11}
\pi_n:\quad S^3/\Z_n\ \stackrel{S^1}{\longrightarrow}\ \CPP^1\ .
\end{equation}
To describe the motion of a point in the cone $(\C^2{\setminus}\{0\})/\Z_n$ over $S^3/\Z_n$, it is sufficient to replace $\vph$ in \eqref{6.8} with $\vph +\omega t$. Note that the exponents in \eqref{6.8} have the opposite signs compared to \eqref{5.12} since the embedding requires changing the orientation on the sphere $S^3$ to match the orientation on $\CO(-1)$. Otherwise, it is the same manifold and the same motion of a point along a circle $S^1$ in fibres of the bundle \eqref{6.11}. To summarize, we reproduce diagram \eqref{1.3} from the introduction
\begin{equation}\label{6.12}
\begin{picture}(100,60)
\put(5.0,0.0){\makebox(0,0)[r]{$\CO(n)^{\times}$}}
\put(55.0,0.0){\makebox(0,0)[c]{${\supset }L(n,\!1){\cong}S^3{/}\Z_n{\subset}\ $}}
\put(100.0,0.0){\makebox(0,0)[l]{$(\C^2{\setminus}\{0\})/\Z_n$}}
\put(5.0,45.0){\makebox(0,0)[r]{``almost quantum"}}
\put(105.0,45.0){\makebox(0,0)[l]{``classical"}}
\put(55.0,45.0){\makebox(0,0)[c]{$(\C^2{\setminus}\{0\})\times\C_{(n)}$}}
\put(5.0,25.0){\makebox(0,0)[c]{$\C^*$}}
\put(110.0,25.0){\makebox(0,0)[c]{$\C^*$}}
\put(25.0,30.0){\vector(-1,-1){18}}
\put(85.0,30.0){\vector(1,-1){18}}
\end{picture}
\\[5pt]
\end{equation}
We wrote ``almost quantum" on the left because we are considering the rotation of a point $\psi_n\exp(-\im\omega nt)$ in the quantum bundle  $\Lv$ and this rotation is equivalent to a $\Z_n$-invariant motion of a point in the phase space $\C^2{\setminus}\{0\}$. On the left and right in \eqref{6.12} we have a description of the same classical $\Z_n$-invariant oscillator as a point rotating in a circle $S^1\subset S^3{/}\Z_n$.

In the spaces of diagram \eqref{6.12} $\psi_n$ is a coordinate of the classical oscillator. The transition to a quantum oscillator is carried out in two steps. First we need to multiply $\psi_n$ by $\psi_0^{\sf v}$, and get a point $\psi_n\psi_0^{\sf v}\exp(-\im\omega (n+1)t)$ with energy $E_n+E_0^{\sf v}=\hbar\omega (n+1)$ rotating in the total space of the bundle $\Lv$. Next, point $(z_0, z_1, \psi_n\psi_0^{\sf v})$ should be replaced by section $(z_0, z_1, \psi_n(z_0, z_1)\psi_0^{\sf v}(z_0, z_1))$, obtaining a sphere $\CPP^1$ with $n$ punctures from \eqref{5.23}, embedded into $\CO(n)\otimes\CO(0)$ and rotating as a whole. When mapped to the orbifold $(\C^2{\setminus}\{0\})/\Z_n$, we obtain an algebraic curve covering this sphere $n$ times. A quantum oscillator in energy state $\tilde E_n=\hbar\omega (n+1)$ is a space $\CPP^1_{(n)}\times S^1$ rotating in the quantum bundle $\Lv$. Sphere $\CPP^1_{(n)}$ specifies the possible initial data for this movement, and function $|\psi_n|^2\exp(-|z|^2)$ is associated with the probability of observing certain initial data.

\smallskip

\noindent {\bf Phase uncertainty.} Let us see how the imposition of $\Z_n$-invariance condition on the motion of a standard classical oscillator looks at the level of formulae \eqref{2.1}-\eqref{2.14}. To begin with, we note that the vector field $V_H$ from \eqref{2.6} is the generator of the group U(1)$_H$ that define the  dynamics \eqref{2.8} of the particle. In spherical coordinates \eqref{2.11} it has the form
\begin{equation}\label{6.13}
V_H=\im\omega z_\al\dpar_{z_\al}^{}-\im\omega \zb_\dal\dpar_{\zb_\dal}^{}=\omega\dpar_\vph\ .
\end{equation}
Accordingly, the Hamiltonian equations of motion have the form
\begin{equation}\label{6.14}
(\dpar_t - V_H)\tilde z_\al =0\quad\Rightarrow\quad(\dpar_t - \omega\dpar_\vph)\tilde z_\al =0\ ,
\end{equation}
where $\tilde z_\al =\psi_{-1}z_\al$ from \eqref{2.11}, and $\psi_{-1}$ depends only on $t$, and $z_\al$ depends on spherical coordinates, but not on time. For $\psi_{-1}=\exp(\im\omega t)$ we get the solution $\tilde z_\al =z_\al (\vph +\omega t)$ from \eqref{2.14} which {\it implicitly assume} that $0\leq\vph <2\pi$. However, this is not a necessary requirement  and the periodicity can be considered with $0\leq\vph <2\pi/n$, $n\in\Nbb$. If we fix $n\ne 1$ then equation \eqref{6.14} should be rewritten as 
\begin{equation}\label{6.15}
(\dpar_t - \omega_n\dpar_{\vph_n}^{})\tilde z_\al =0 \for \omega_n=n\omega\ ,\ \ 0\leq\vph_n=n\vph<2\pi\ ,
\end{equation}
where the redefinition $\vph\mapsto\vph_n$ is necessary for the standard definition of periodicity. Accordingly, the solution to the equation \eqref{6.15} will be
\begin{equation}\label{6.16}
(\tilde z_\al)=\left(\psi_{-n}^{\frac{1}{n}}z_\al \bigl(\frac{\vph_n}{n}\bigr)\right)=\left[\rho e^{\frac{\im}{n}(\vph_n+\omega_nt)}e^{-\frac{\im\chi}{2}}\right]\frac{(1,z)}{\sqrt{1+z\zb}}
\end{equation}
for the patch $U_0$ and similarly for the patch $U_1$ on $\CPP^1$. It follows from \eqref{6.15} that the period of motion of a point along a circle in \eqref{6.16}  is equal to $T_n=2\pi/\omega_n$, and the coordinates $z_\al$ are subject to the equivalence relation \eqref{6.6}. Only the domain of definition of the coordinate $\vph$ in the spherical coordinate system on $\C^2{\setminus}\{0\}$ changes.

In the sense described above, the angular coordinate $\vph$ is not completely defined since it can have an infinite number of different domains of definition $[0, 2\pi/n)$ with $n=1,2,3,...\,$. This uncertainty leads to the fact that the same equation \eqref{6.14} describes the motion \eqref{6.16} of a classical oscillator in different subspaces $(\C^2{\setminus}\{0\})/\Z_n$ of the phase space $\C^2{\setminus}\{0\}$. Formula \eqref{6.16} defines the $\Z_n$-invariant motion of the classical oscillator. The local coordinates \eqref{6.16} define a $\Z_n$-invariant motion on the right side of diagram \eqref{6.12} and it is equivalent to a $\Z_n$-invariant motion on the left side of this diagram.

Note that for $n\geq 2$ the variables $z_\al$ from \eqref{6.16} cease to be $\Z_n$-invariant variables, and we are forced to explicitly break this symmetry by fixing the domain $0\leq\vph <2\pi/n$. Invariant variables on $(\C^2{\setminus}\{0\})/\Z_n$ for $n>1$ are $n+1$ functions $\{\psi_{nm}(z), m=0,...,n\}$ from  
 \eqref{4.10} generating the ring of invariants on the space $(\C^2{\setminus}\{0\})/\Z_n$. It is these functions that are combined into the ``wave function" $\psi_n(z_0,z_1)\exp(-\im\omega_nt)$ from \eqref{4.3}-\eqref{4.5}, which is vanished by the differential operator in equation \eqref{6.15}. In other words, the holomorphic isomorphism from $(\C^2{\setminus}\{0\})/\Z_n$ to $\CO(n)^\times$ in diagram \eqref{6.12} defines a transition from non-invariant to invariant variables.

\smallskip

\noindent {\bf States $\psi_{nm}$.} We discussed the eigenfunctions $\psi_n(z_0, z_1)$ of the Hamiltonian $\hat H$ that parametrize the subspace $\C^{n+1}$ of the Hilbert space $\CH$. They correspond to $\Z_n$-invariant coordinates of the fibres of the bundles $\CO(n)$ over $\CPP^1$ and define conical spaces $(\C^2{\setminus}\{0\})/\Z_n$ with $n\in\Nbb$ according the diagram \eqref{6.12}. Note that the basis in the Hilbert space $\CH$ is formed not by functions $\psi_n$, but by functions $\psi_{nm}(z_0, z_1)$ obtained from $\psi_n$ by setting to zero all complex numbers $b_{n\ell}$ with $\ell\ne m$. Therefore, these functions $\psi_{nm}$ inherit the description of the states  $\psi_n$ and have additional symmetries, which we will now describe.

The bundles $\CO(n)$ are defined by divisors \eqref{5.22} and sections $\psi_{nm}(z_0, z_1)$ correspond to divisors of the form
\begin{equation}\label{6.17}
D_{nm}=mP_0+(n-m)P_\infty\ ,
\end{equation}
where $P_0{=}[1:0]\sim z{=}0$ and $P_\infty{=}[0:1]\sim w{=}0$. Let us define the action of the group $\C^*$ on $\CPP^1$ by the formula $\lambda_1\cdot[z_0:z_1]=[z_0:\lambda_1z_1]$. Points $P_0$ and $P_\infty$ are fixed under this action, i.e. the divisor \eqref{6.17} is invariant. The divisor \eqref{6.17} corresponds to a bundle $\CO_m(n)$, which is called $\C^*$-equivariant. This means that its total space is subject to an action of the group $\C^*$ that does not change its structure. In the case of bundle $\CO_m(n)$, this is the action of the group $\C^*$ with weight $w_0=-m$ and $w_\infty=n-m$ of fibres over the points $P_0$ and $P_\infty$, so that $w_\infty -w_0=n$. The number $m$ in $\CO_m(n)$ characterizes the distinctness of the equivariant bundle $\CO_m(n)$ compared to bundles $\CO(n)$, where the divisor \eqref{5.22} is not invariant under the action of the group $\C^*$ of the type described above. Otherwise, everything said about diagram \eqref{6.12} remains the same, just $\C_{(n)}$ is replaced by $\C_{(nm)}$ and $\psi_n$ is replaced by $\psi_{nm}$. The sections of the bundle $\CO_m(n)$ will be the functions \eqref{4.10}, and not their sum over $m$.

It should be noted that the subbundle of unit vectors of the equivariant bundle  $\CO_m(n)$ will not simply be a lens space  \eqref{5.12}, but a Seifert fibration $L(n,m;1)$, whose characteristics contain information about the action of the subgroup U(1) of the group $\C^*$ preserving the divisor \eqref{6.17}. Accordingly, on the right-hand side of the diagram \eqref{6.12}, we obtain a Seifert fibration, not a lens space. Moreover, all these spaces have the same topology, since $m$ is a non-topological group-theoretic characteristic, so summation over $m$ is possible.

\section{Ground state and zero-point energy}

\noindent {\bf Ground state.} Our consideration so far has been incomplete. Namely, from the diagram \eqref{1.3} to diagram \eqref{6.12}, we excluded from consideration the case $n=0$ and the point $\{0\}$ from the phase space $\C^2$. In diagram \eqref{1.4}, we claimed that both the bundle $\CO(0)=\CPP^1\times \C_{(0)}$ and the point $\{0\}\in\C^2$ are associated with the ground state of the oscillator. We will now examine this correspondence in detail, both at the level of the bundle coordinates and at the level of sections.

\smallskip

\noindent {\bf Blow-up of $\C^2$.} Let us consider a sphere $\CPP^1$ with homogeneous coordinates $[z_0:z_1]$. A point on the sphere $\CPP^1$ defined by these coordinates can be associated with a non-zero vector $(z_0, z_1)\in\C^2$ from point $\{0\}$ to $\{z_\al\}\in\C^2$. This vector defines a line in $\C^2$ whose points have coordinates $\psi_{-1}z_\al$, where $\psi_{-1}$ is a complex number. Thus a point $[z_0:z_1]$ on the sphere $\CPP^1$ defines the direction of a vector in the space $\C^2$, $\psi_{-1}$ defines its length, and the set of points
\begin{equation}\label{7.1}
\bigl([z_0:z_1], \psi_{-1}z_{\al}\bigr)\in \CPP^1\times\C^2
\end{equation}
defines a tautological line bundle over $\CPP^1$ denoted $\CO(-1)$ (see e.g. \cite{Wells}). Here $-1$ is the Chern class of the holomorphic line bundle $\CO(-1)$. 

Using definition \eqref{7.1}  we can define projections 
\begin{equation}\label{7.2}
\begin{picture}(100,60)
\put(70.0,0.0){\makebox(0,0)[l]{$\C^2\ni\tilde z_\al =\psi_{-1}z_\al$}}
\put(10.0,0.0){\makebox(0,0)[r]{$[z_0{:}z_1]\in\CPP^1$}}
\put(40.0,45.0){\makebox(0,0)[c]{$\CO(-1)$}}
\put(10.0,45.0){\makebox(0,0)[r]{$([z_0{:}z_1], (\tilde z_0,\tilde z_1))\in$}}
\put(0.0,25.0){\makebox(0,0)[c]{$\pi$}}
\put(75.0,25.0){\makebox(0,0)[c]{$f$}}
\put(20.0,35.0){\vector(-1,-1){25}}
\put(50.0,35.0){\vector(1,-1){25}}
\end{picture}
\\[10pt]
\end{equation}
where we already introduced the rescaled coordinates $\tilde z_\al$ in \eqref{2.11}, and the projection onto $\CPP^1$ in \eqref{2.13}. If $\psi_{-1}\ne 0$ then $f$ is the identity mapping of $\CO(-1)^\times$ into $\C^2{\setminus}\{0\}$, i.e. $\CO(-1)^\times\cong\C^2{\setminus}\{0\}$ as in \eqref{2.11}-\eqref{2.13}. When $\psi_{-1}=0$, we obtain the point $\tilde z_\al=0$ from the mapping $f$, and the preimage of this point is the sphere $f^{-1}(\{0\})=\CPP^1\subset\CO(-1)$. This sphere is zero section of the bundle $\pi :  \CO(-1)\to\CPP^1$ on the left side of diagram \eqref{7.2}. This is why $\tilde\C^2:=\CO(-1)\cong\CPP^1\sqcup(\C^2{\setminus}\{0\})$ is called the blow-up of the space $\C^2=\{0\}\sqcup(\C^2{\setminus}\{0\})$ at the point $\{0\}$, which is replaced by the sphere $\CPP^1$. In algebraic geometry, a blow-up is a geometric transformation used to unfold or resolve singularities (points where a shape crosses itself or has a sharp cusp).

\smallskip

\noindent {\bf Ground state diagram.} Using the correspondence \eqref{7.2} we can introduce an analogue of diagram \eqref{6.12} for the case $n=0$. In this case of a trivial bundle $\CO(0)=\CPP^1\times\C_{(0)}$ with zero Chern class $n=0$, one should simply define the direct product of all spaces in \eqref{7.2} with the fibre $\C_{(0)}$ of the bundle $\Lv$, obtaining
\begin{equation}\label{7.3}
\begin{picture}(100,60)
\put(70.0,0.0){\makebox(0,0)[l]{$\C^2\times\C_{(0)}$}}
\put(10.0,0.0){\makebox(0,0)[r]{$\CPP^1\times\C_{(0)}$}}
\put(40.0,45.0){\makebox(0,0)[c]{$\tilde\C^2\times\C_{(0)}$}}
\put(5.0,45.0){\makebox(0,0)[r]{``almost quantum"}}
\put(75.0,45.0){\makebox(0,0)[l]{``classical"}}
\put(0.0,25.0){\makebox(0,0)[c]{$\pi$}}
\put(75.0,25.0){\makebox(0,0)[c]{$f$}}
\put(20.0,35.0){\vector(-1,-1){25}}
\put(50.0,35.0){\vector(1,-1){25}}
\end{picture}
\\[10pt]
\end{equation}
We write ``almost quantum" since we are considering points and not section, but still quantum since these points move in fibres $\C_{(0)}$ of the quantum bundle $\Lv$. To do this, it is sufficient to replace the coordinates $z_\al$ in the function $\psi_0^{\sf v}(z,\zb, \theta)$ from \eqref{3.5} with the parameters $y_\al$, after which $\psi_0^{\sf v}$  will simply be a complex number from the point of view of phase space $\C^2$. The space $\C_{(0)}$ is parametrized by the coordinate $\psi_0^{\sf v}(t)$ and below we will show that it depends on time as $\psi_0^{\sf v}(t)=\psi_0^{\sf v}\exp(-\im\omega t)$, and the coordinates $\tilde z_\al\in\C^2$ do not depend on time.

Diagram \eqref{7.3} states that the particle corresponding to the ground state moves along a circle $S^1$ with $\theta(t)=\theta-\omega t$ in the internal space $\C_{(0)}$ and is at rest at an arbitrary point $\tilde z_\al$ in the phase space $\C^2$. It is reasonable to consider it as resting at the origin $\{0\}\in\C^2$ as in diagram \eqref{1.4} from the Introduction.  If we replace the point $\psi_0^{\sf v}\in\C_{(0)}$ with a section, then from the norm $|\psi_0^{\sf v}|^2=\exp(-|z|^2)$  of the section there will follow a Gaussian smearing around the point $\{0\}\in\C^2$. 

Together, diagrams \eqref{6.12} and \eqref{7.3} define the coordinate of the fibre in $\Lv$ as $\psi_n\psi_0^{\sf v}$, where the part $\psi_n$ defines the motion in the subspace $(\C^2{\setminus}\{0\})/\Z_n$ of the phase space $\C^2$, and the ground state $\psi_0^{\sf v}$ always lies outside the classical phase space $\C^2$. Thus, the quantum state is not reducible to a classical one due to the purely quantum nature of the ground state.

\smallskip

\noindent {\bf Schr\"odinger equation.} It is useful to represent the fibre coordinates $\psi_n$ of the bundle $\Lv$ as functions \eqref{5.5} of the auxiliary coordinates $y_\al$, and similarly for the function $\psi_0^{\sf v}$ from \eqref{3.5}.
Then the transition from fibre coordinates to sections is accomplished simply by replacing $y_\al$ with $z_\al$, and from sections to coordinates by replacing $z_\al$ with $y_\al$.

To understand why we have constant rotation in the fibre of the quantum bundle $\Lv$, let us consider the Schr\"odinger equation for a quantum oscillator more closely. In general, the covariant Laplacian \eqref{4.1} should be written as follows (see e.g. Sect.3.2 in \cite{Popov1})
\begin{equation}\label{7.4}
\hat H=-\sfrac12\hbar\omega\Delta_2=-\hbar\omega\delta_{\dal\al}\nabla_{\zb_\dal}^{}\nabla_{z_\al}^{}-\sfrac12\hbar\omega\delta_{\al\dal}F^{\al\dal}\ ,
\end{equation}
where
\begin{equation}\label{7.5}
\nabla_{z_\al}^{}=\dpar_{z_\al}^{}+A_{z_\al}^{}\dpar_\theta=\dpar_{z_\al}^{}+\sfrac12\delta^{\al\dal}\zb_\dal\dpar_\theta\ ,\quad
\nabla_{\zb_\dal}^{}=\dpar_{\zb_\dal}^{}+A_{\zb_\dal}^{}\dpar_\theta=\dpar_{\zb_\dal}^{}-\sfrac12\delta^{\dal\al}z_\al\dpar_\theta\ ,
\end{equation}
\begin{equation}\label{7.6}
F^{\al\dal}=\left[\nabla_{z_\al}^{},  \nabla_{\zb_\dal}^{}\right]=\im\delta^{\al\dal}\dpar_\theta\with \psi_0^{\sf v}=|\psi_0^{\sf v}|\exp(\im\theta)\ .
\end{equation}
Operator
\begin{equation}\label{7.7}
-\im\dpar_\theta=\psi_0^{\sf v}\dpar_{\psi_0^{\sf v}}^{}-\bar\psi_0^{\sf v}\dpar_{\bar\psi_0^{\sf v}}^{}
\end{equation}
in formulae \eqref{7.4}-\eqref{7.7} is usually replaced by 1, thereby losing information about $\theta{=}$arg$\psi_0^{\sf v}$. If this is not done, then the Schr\"odinger equation \eqref{4.4} takes the form
\begin{equation}\label{7.8}
\begin{split}
&\im\,\hbar\,\dpar_t(\psi_n\psi_0^{\sf v})=\hat H(\psi_n\psi_0^{\sf v})=|\psi_0^{\sf v}|\hbar\omega(z_\al\dpar_{z_\al}^{}-\im\dpar_\theta)(\psi_n e^{\im\theta})\ ,
\\[2pt]
&\Rightarrow\hbar\omega (z_\al\dpar_{z_\al}^{}-\im\dpar_\theta)(\psi_n e^{\im\theta})=(E_n+E_0^{\sf v})(\psi_n e^{\im\theta}) \ ,
\end{split}
\end{equation}
where $\psi_n$ depends on $t$ through $\exp(-\sfrac\im\hbar\,E_n t)$ and $\psi_0^{\sf v}$ through $\exp(-\sfrac\im\hbar\,E_0^{\sf v}t)$. We obtain solution
\begin{equation}\label{7.9}
\Psi(t)=\left(\sum_{n=0}^\infty c_n\psi_n(z)e^{-\im\omega nt}\right)(\psi_0^{\sf v}e^{-\im\omega t})\ ,
\end{equation}
where the last factor in \eqref{7.9} defines the Hermitian metric \eqref{3.6} on fibres and the Gaussian measure in the inner product \eqref{3.9}, as well as the rotation in the internal space $\C_{(0)}$.

\smallskip

\noindent {\bf Oscillations in internal space.} For the coordinate on the fibre $\C_{(0)}$ of the bundle $\Lv$ over $\C^2$ we have
\begin{equation}\label{7.10}
\dpar_t\psi_0^{\sf v}=-\omega\dpar_\theta\psi_0^{\sf v}=-\im\omega\psi_0^{\sf v}\ ,
\end{equation}
which corresponds to a rotation around a circle $S^1$ in the phase space $\C_{(0)}$ or an oscillation on the real line Re$\psi_0^{\sf v}$. In the ground state, the point moving in $\Lv$ is at rest at $\{0\}\in\C^2$, and rotation occurs only in the fibre $\C_{(0)}$ of the bundle $\Lv$ with a constant angular velocity $\omega$. This velocity is given by the last term in \eqref{7.4} with the curvature $F^{\al\dal}$ of the bundle $\Lv$ convoluted with the symplectic 2-form $\Omega_{\al\dal}$ of the phase space $\C^2$. The energy of this state is equal $E_0^{\sf v}=\hbar\omega$. It is the presence of this rotational energy in fibres that leads to a non-zero curvature $\Fv$ in \eqref{7.6} of the quantum bundle $\Lv$.

Thus, the ground-state energy {\it curves the total space} of the bundle $\Lv$ over the {\it phase space} $\C^2$, but {\it does not curve the base} $\C^2$ of this bundle. This explains why the enormous vacuum energy does not curve space-time. Recall that the curvature $\Fv$ of the quantum bundle is zero along coordinate space, unlike, for example, the Maxwell field strength. The internal phase space $\C$ of a particle really exists, and a quantum particle differs from a classical particle in that it moves not only in the ordinary phase space of coordinates and momenta, but also along a circle $S^1$ within the fibres $\C$ of quantum bundle. The energy of this motion is the energy of the ground state, which determine the curvature of the quantum bundle. The particles are in the fundamental representation  $\exp(\im\theta)$ of the structure group $\sU(1)_{\sf v}$ of the quantum bundle, and the antiparticles are in the conjugate representation $\exp(-\im\theta)$ of the group $\sU(1)_{\sf v}$ \cite{Popov2}.

\section{Background field $\Av$ and interactions}

\noindent {\bf Extended phase space $\Lv$.} Let us  sum up the preliminary results. In Sections 1-4 we recalled the standard descriptions of the classical and quantum harmonic oscillator. In the quantum case, we emphasize the fact that quantum mechanics is a gauge theory on a phase space with a fixed connection $\Av$ and curvature $\Fv$ \cite{Sni, Wood, Hurt}. The eigenfunctions of the quantum Hamiltonian have the form
\begin{equation}\label{8.1}
\Psi_n(z)=\psi_n(z)\psi_0^{\sf v}(z)\ ,
\end{equation}
where the explicit form of the polynomials $\psi_n(z)$ and the ground state $\psi_0^{\sf v}(z)$ is given in \eqref{3.5}, \eqref{3.6} and \eqref{4.3}. The polynomial part $\psi_n(z)$ of the wave function \eqref{8.1} is associated with the energy $E_n=\hbar\omega n$, and the ground state $\psi_0^{\sf v}(z)$ is associated with the energy $E_0=\hbar\omega$.

The main idea of the paper is to move from sections $\Psi_n(z)$ of the bundle $\Lv$ described in Sections 3 and 4 to coordinates \eqref{3.2} on it as on an extended phase space. We called the particle moving in this space $\Lv$ an almost quantum oscillator. We considered  $\psi_n(z)$ and $\psi_0^{\sf v}(z)$ separately, since they are fundamentally different entities. To go from function $\psi_0^{\sf v}(z)$ to a coordinate on a fibre, $\psi_0^{\sf v}$ should be considered as a complex number (scalar). In the functions $\psi_n(z)$, it is convenient to replace $z_\al$ with external parameters $y_\al$ of weight +1 with respect to the group $\C^*$, which allows one to clearly see and control the fact that the coordinates $\psi_n$ of the form \eqref{5.5} have weight $n$ under the action of the group $\C^*$ (equivariance of the bundle $Y_n$). In Section 5 and 6 we showed that the coordinate $\psi_n$ defines the motion of a classical $\Z_
n$-invariant oscillator in the lens space $S^3/\Z_n$ embedded in biholomorphic spaces $(\C^2{\setminus}\{0\})/\Z_n\cong\CO(n)^\times$. This oscillator moves with period $T_n=2\pi/\omega_n$ and angular velocity $\omega_n=n\omega$ along a circle $S^1$ in $S^3/\Z_n$.

The ground state $\psi_0^{\sf v}$ corresponds to an almost quantum oscillator moving with a constant angular velocity $\omega$ along a circle $S^1$ in the fibre $\C_{(0)}$ of the bundle $\Lv$ over $\C^2$ and at rest at the point $\{0\}$ of the phase space $\C^2$. Here $\C_{(0)}$ with coordinate $\psi_0^{\sf v}$ is the space of fundamental representation of the structure group $\sU(1)_{\sf v}$ of the bundle $\Lv$. The meaning of this statement is revealed by formulae 
\eqref{7.4}-\eqref{7.9}, where $\dpar_\theta$ in \eqref{7.5}-\eqref{7.8} is the generator of the group U(1)$_{\sf v}$, and $\psi_0^{\sf v}=|\psi_0^{\sf v}|\exp(\im\theta)$. Thus, the almost quantum oscillator given by the product $\psi_n\psi_0^{\sf v}$ describes a particle moving along a circle $S^1$ in $S^1_{\sf cl}\times S^1_{\sf qu}$ with $S^1_{\sf cl}\subset\C^2$, $S^1_{\sf qu}\subset\C_{(0)}$ in $\Lv\cong\C^2\times\C_{(0)}$.

\smallskip

\noindent {\bf Interaction with $\Av$.} Note that the dynamics of an almost quantum oscillator is of the form
\begin{equation}\label{8.2}
\Psi_n(t)=\left(\psi_ne^{-\im\omega nt}\right)  \left(\psi_0^{\sf v}    e^{-\im\omega t}\right)  \ ,
\end{equation}
where $\Psi_n$ is a coordinate in the tensor product of the bundles $\CO(n)_H\otimes\CO(0)_{\sf v}$ over $\CPP^1$ lifted to $\C^2$. Here $z_\al\in\C^2$, $\psi_n\in \C$ and $\psi_0^{\sf v}\in\C_{(0)}$ are the initial data for solutions \eqref{5.16}, \eqref{6.16} and \eqref{7.10} glued in \eqref{8.2}. Superposition of solutions with different values of $n$ is impossible for either a classical or an almost quantum oscillator. Such a superposition is possible when passing from the coordinates of the space $\Lv$ to sections, when $n$ and $m$ cease to be bundle invariants and simply define the grading of the Hilbert space $\CH$.

Recall that the quantum bundle $\Lv\to\C^2$ is a complex vector bundle of rank one, so $\psi_0^{\sf v}$ is a complex vector. We want to move this vector from the point $\{0\}\in\C^2$ to an arbitrary point $\{z_\al\}\in\C^2$. The complex geometry of the bundle $\Lv$ with a connection $\Av$ imposes strict restrictions on these shifts. The shift of fibres $\C$ must preserve polarization, in this case holomorphicity, which is expressed by equation \eqref{3.5} for the shifted vector $\psi_0^{\sf v}(z)$. Accordingly, the general section has the form  \eqref{3.6}. Thus, the wave functions \eqref{8.1} contain information not only about the initial data of the motion of an almost quantum oscillator in space $\Lv$, but also the reaction of geometry to a change in the initial data $z_\al\sim (x_\al , p_\al)$. And since we consider sections \eqref{3.9} normed on unity, they can be given a probabilistic character associated with the Hermitian metric \eqref{3.6} on fibres. The greater the shift in $z_\al\in\C^2$, the stronger the resistance of curvature $\Fv=\dd\Av$ and the lower the probability of the system being in the vicinity of the shifted point $z_\al$. We emphasize that a shift of point $z_\al$ implies a change in both coordinates and momenta (initial data), and therefore a change in energy. This is possible only due to the action of an external force on a particle moving in space $\Lv$.

\smallskip

\noindent {\bf Change of states.} The background connection $\Av$ is responsible not only for specifying the explicit form \eqref{3.6} and \eqref{8.1} of holomorphic sections, but also for changing the degree $n$ of the polynomials $\psi_n(z)$ in \eqref{8.1}, that is, for mapping different states into each other. For sections \eqref{8.1}, the action of the covariant derivatives \eqref{7.5} on $\Psi_n$ has the form
\begin{equation}\label{8.3}
\nabla_{z_\al}^{}(\psi_n\psi_0^{\sf v})=\left(\dpar_{z_\al}^{}\psi_n\right)\psi_0^{\sf v}\und
\nabla_{\zb_\dal}^{}(\psi_n\psi_0^{\sf v})=-\delta^{\dal\al}z_\al\psi_n\psi_0^{\sf v}\ ,
\end{equation}
that is the operator $\nabla_{z_\al}^{}$ maps $\Psi_n$ into $\Psi_{n-1}$, and the operator $\nabla_{\zb_\dal}^{}$ maps $\Psi_n$ into $\Psi_{n+1}$. Thus, change in states $\Psi_n\mapsto\Psi_{n\pm 1}$ occur due to their interaction with the background gauge field $\Av$. The strength of this field $\Fv$ is constant and we have an analogue of a charged particle in a constant magnetic field with a quantum charge $\qv=1$ instead of electric charge. Unlike the electromagnetic field, which is defined in coordinate space, fields $\Av$ and $\Fv$ are defined in phase space.

When passing to purely holomorphic functions $\psi_n(z)$ and transferring $\psi_0^{\sf v}$ to the integration measure \eqref{3.9}, the covariant derivative $\nabla_{z_\al}^{}$ and $\nabla_{\zb_\dal}^{}$ are transformed into operators of differentiation $\dpar_{z_\al}^{}$ and multiplication by $z_\al$, and the memory  of the interaction with the field $\Av$ is erased. Then, instead of the standard description used in differential geometry and gauge theories, one begins to talk about mysterious ``quantum operators".

\smallskip

\noindent {\bf Coherent states.} Coherent states can serve as an illustration of what has been said above. Consider the operator
\begin{equation}\label{8.4}
\exp(-\nabla(b))\with \nabla(b)=b_\al\nabla_{z_\al}^{}+\bar b_\dal\nabla_{\zb_\dal}^{}\ ,
\end{equation}
where $b_\al$ are complex numbers and $\bar b_\dal=(b_\al)^*$. Let us apply this operator to the ground state $\psi_0^{\sf v}$,
\begin{equation}\label{8.5}
\exp(-\nabla(b))\psi_0^{\sf v}=(D(b)\cdot 1)\psi_0^{\sf v}= e^{-\frac12\bar b^\al b_\al}   e^{\bar b^\al z_\al}  \psi_0^{\sf v}=:\Psi(b)\psi_0^{\sf v}    \ ,
\end{equation}
where
\begin{equation}\label{8.6}
D(b)= e^{\bar b^\al a^\+_\al -b_\al a^\al} \with   a^\+_\al =z_\al\und a^\al = \dpar_{z_\al}^{}  
\end{equation}
is a unitary displacement operator, $\bar b^\al =\delta^{\al\dal}\bar b_\dal$. Thus, a coherent state $\Psi(b)$ in \eqref{8.5} arises when a particle in the ground state interacts with the background gauge field $\Av$ entering in the covariant derivatives in \eqref{8.4}.

\smallskip

\noindent {\bf Reducible representations.} Superposition of $\Z_n$-invariant solutions for a classical oscillator with different $n$ on the spaces $(\C^2\setminus\{0\}/\Z_n\subset\C^2$ and $\CO(n)^\times$ is impossible for topological reason. However, such superposition becomes possible after lifting to the extended phase space $\Lv$, where the numbers $n$ and $m$ cease to be invariants and become simply grading parameters. The general solution of the Schr\"odinger equation is a superposition of states \eqref{8.1} with complex coefficients $c_n$, giving a vector $\Psi$ of the Hilbert space $\CH$,
\begin{equation}\label{8.7}
\Psi=\sum_{n=0}^\infty c_n\psi_n\psi_0^{\sf v}\in \CH =\mathop\oplus_{n=0}^\infty \C^{n+1}\otimes\C_{(0)}\ ,\quad
\C^{n+1}=\mathop\oplus_{m=0}^n\C_{(nm)}\ .
\end{equation}
Here $\C_{(0)}$ is the space of fundamental representation of the group U(1)$_{\sf v}$, and $\C^{n+1}$ is the space of irreducible representation of the group SU(2) acting on the phase space $\C^2$. On the space $\C_{(nm)}$, an irreducible representation of weight $n$ of the group U(1)$_{H}$ and an irreducible representation of weight $n-2m$ of the subgroup U(1)$_S$ of the group SU(2) are defined.

The Hilbert space $\CH$ itself is a {\it reducible} representation of all these groups. It is precisely the reducibility of the representation of the group U(1)$_{H}$ and SU(2) on the space $\CH$ that is responsible for the fact that the vector $\Psi$ from \eqref{8.7} is not an eigenvector for the energy operator $\hat H$ and for the generators of the group SU(2). We call the state $\Psi$, which is a superposition of all classical states $\psi_n$ supplemented by the ground state $\psi_0^{\sf v}$, a quantum particle. It is impossible to assign fixed numbers $(n,m)$ to a quantum particle until the function $\Psi$  coincides with $\psi_{nm}\psi_0^{\sf v}\subset\C_{(nm)}\otimes\C_{(0)}\subset\CH$.

We obtain the following correspondence between observable (classical) and unobservable (quantum) states:
\begin{equation}\nonumber
\begin{array}{lc}
\mbox{``classical"} &\mbox{``quantum"}   \\[1pt]
(z_\al)\in\C^2&(z_\al, \Psi)\in\C^2\times\C\\[1pt]
\mbox{irreducible representation}\quad\quad&\mbox{infinite-dimensional reducible representation}\\[2pt]
\psi_n\in\C^{n+1}, \psi_{nm}\in\C_{(nm)}&\Psi=\sum\limits_{n=0}^\infty c_n\psi_n\psi_0^{\sf v}\in\mathop\oplus\limits_{n=0}^\infty\C^{n+1}\otimes\C_{(0)}
\end{array}
\end{equation}
Quantum effects arise at the level of reducible representations. Vectors from different subspaces $\C^{n+1}$ have different phases, specified by the parameter $n$, and interference is possible. The quantum numbers $n$ and $m$  can be ``measured" only by returning from the extended phase space $\Lv\cong\C^3$ to the phase space $\C^2$, that is only through projection onto the subspace of irreducible representation. The Hilbert space $\CH$ itself is necessary to define mapping $\psi_{nm}\mapsto\psi_{n'm'}$ that change quantum numbers due to interaction with the background gauge field $\Av$, as can be seen in formulae \eqref{8.3}. The transition from  classical to quantum level corresponds to the transition from irreducible to reducible representations of the symmetry groups of the system under consideration.

\smallskip

\noindent {\bf Probabilities.} The possibility of a probabilistic Born interpretation is related to the normalization of cross sections and the requirement that polarization be preserved when shifting the fibres of the complex vector bundle along the base manifold. This is true for any complex vector bundle in any dimension and is not related to the ``incomprehensible quantum nature" of our world. In addition, the wave function \eqref{8.7} may reflect the probabilities $|c_n|^2$ of particle motion in one of the subspace $\C^2/\Z_n$ of $\C^2$ if the initial state and its symmetries are not known with absolute accuracy in accordance with 't Hooft idea \cite{Hooft}. Let us also imagine that a particle with fixed parameters $(n,m)$ is lifted into bundle $\Lv$. Then it is acted upon by operators defined as functions of covariant derivatives (like operators \eqref{8.4}), and the particle repeatedly transitions to states with other quantum numbers. Its final state can only be predicted probabilistically, since it is impossible to control all of its interactions, but the process of changing states itself appears deterministic.

\smallskip

\noindent {\bf Interactions with $A_E$.} A section of any complex vector bundle $E$ over any manifold always has an associated charge $q\in\R$ (particles), and a section of its complex conjugate bundle $\bar E$ has an associated charge $-q$ (antiparticles). In the case under consideration, we have a bundle $\Lv$ over the phase space $\C^2$ with a background gauge field $\Av$ entering in the covariant derivatives  \eqref{7.5} and sections $\Psi$ with charge $\qv=1$. It is in the interaction of $\Psi$ with $\Av$ that ``quantum" effects appear. We can endow $\Psi$ with other charges, for example, electric or color. To do this, we must introduce the tensor product of bundles $\Lv\otimes E$, where $E$ is one of the complex vector bundles of the standard model with gauge group SU(3)$\times$SU(2)$\times$U(1). Then the corresponding gauge field $A_E$ will enter the covariant derivatives acting on the sections $\Psi_E$ of the bundle $\Lv\otimes E$. The difference between the gauge field $A_E$ of the standard model and $\Av$ is that they are dynamic and their curvature is non-zero only along the coordinate subspace of the phase space. For curvature $\Fv$, only components mixed in coordinates and momenta are non-zero. In mathematical literature, electromagnetic fields $A_{\sf em}$, $F_{\sf em}$ are considered as deformations of fields $\Av, \Fv$, which are always non-zero.

\smallskip

\noindent {\bf Measurement of observables.}  At a fundamental level, changes of states are always caused by interactions with gauge fields -- carriers of interactions -- on the complex vector bundle $\Lv\otimes E$. The point is that covariant derivatives are generalized momenta and their effect on sections $\Psi_E$ of the bundle $\Lv\otimes E$ changes the momentum of the particle. For example, right- and left-handed photons in the momentum representation are complexified light-like momenta and the exchange of photons is an exchange of momenta (see e.g. \cite{Popov1}). In order to ``measure" the parameters of a state, one must act on it with an operator constructed from covariant derivatives (``send a photon"), and this is always an action in the quantum domain, and this action changes the momentum and energy of the state.

In the measurement process, not only the point $z_\al\sim (x_\al , p_\al)$, to which the fibre $\C$ of the quantum bundle is attached, is always involved, but also the shift $\Delta z_\al\sim(\Delta x_\al, \Delta p_\al)$ of the point of attachment of $\C$ to $\C^2$. This shift is given by covariant derivatives in the bundle $\Lv$ over $\C^2$. The commutators of these covariant derivatives is the curvature $\Fv$ of the bundle and it is this that defines the Heisenberg uncertainty relations $\Delta x_\al\Delta p_\beta\geq\sfrac12\hbar\delta_{\al\beta}$. Thus, the probabilistic interpretation and the uncertainty relation are not connected with the coordinates and momenta $(x_\al , p_\al)$ themselves, but with the reaction of the curvature $\Fv$ to their change  $(\Delta x_\al , \Delta p_\al)$, which in turn is connected with the ``measurement".

At a fundamental level, measurement changes state, and one should speak of covariant derivatives with gauge fields $\Av$ and $A_E$ on vector bundles $\Lv\otimes E$ instead of mathematically vague concepts such as ``device", ``environment" and ``observer", in terms of which the discussion is usually conducted (see e.g. \cite{S1, Tomaz}). At the fundamental level, there is no boundary between quantum and classical, since covariant derivatives on the bundle $\Lv\otimes E$ are ``quantum" operators. Changes in classical states occur at the quantum level.

\section{Conclusions}

\noindent {\bf Almost quantum and quantum.} In this paper we have shown that the eigenfunctions $\psi_n(z)$ of quantum oscillator correspond to the coordinates of a classical oscillator moving in space $\C^2/\Z_n$ which can be embedded into the phase space $\C^2$. Note that two such oscillators associated with group $\sSU_L(2)\times\sSU_R(2)$ play a key role in deriving the relativistic equations in the geometric reformulation of Wigner's approach \cite{Popov1}. The ground state $\psi_0^{\sf v}$ corresponds to the trivial representation of all groups acting on the phase space and defines an almost quantum oscillator. The coordinate  $\psi_0^{\sf v}$ belongs to the fundamental representation of the structure group U(1)$_{\sf v}$ of the bundle $\Lv$ and describe the constant rotation of the particle in fibres of this bundle.

Quantum properties arise after lifting $\psi_n$ into the extended phase space $\Lv$ and replacing the fibre coordinates with sections. The interaction of sections of the bundle $\Lv$ with the background gauge field $\Av$ allows us to map initial functions that parametrize irreducible representations of the groups U(1)$_H$ and SU(2) into those that parametrize other irreducible representations of these groups and to consider their superpositions.

\smallskip

\noindent {\bf Kepler problem.} Note that exactly the same connection between classical and quantum states as described in this paper also exists for the states of the hydrogen atom. From the point of view of classical mechanics, this is Kepler's problem of motion in a central force field. The phase space of a particle is $T^*\R^3\cong\R^6$ and the motion is given by the Hamiltonian $H=p^2/2m - \mu/r$, where $r^2=x_1^2+x_2^2+x_3^2$ and $\mu>0$. For orbits with constant negative energy $H=E<0$ in the limit $r\to 0$ the momenta become infinite. Two different regularizations of this singularity were proposed by Kustaanheimo and Stiefel (KS) \cite{Stiefel} and Moser \cite{Moser}.

In KS approach, coordinates and time are redefined and the phase space $T^*\R^3$ is embedded into the phase space $T^*\R^4\cong\C^4$ of the harmonic oscillator in four dimensions. The surface of constant energy of this oscillator is the sphere $S^7\subset\C^4$, and the trajectories are parametrized by the complex projective space $\CPP^3=S^7/\sU(1)$. The conserved quantities of the Kepler problem (energy, angular momentum, and Runge-Lenz vector) impose a constraints on the $4D$ oscillator, distinguishing in the sphere $S^7$ a submanifold
\begin{equation}\label{9.1}
T^{1,1}=\sSO(4)/\sSO(2)\cong\sSU(2)\times\sSU(2)/\sU(1)\ \stackrel{S^1}{\longrightarrow}\ \CPP^1\times\CPP^1\ ,
\end{equation}
fibred over the product of two spheres, where the group U(1) is embedded in the group SU(2)$\times$SU(2) as a diagonal subgroup. Topologically, $T^{1,1}$ is isomorphic to the space $S^3{\times}S^2$  and it is this 5-dimensional manifold that defines the regularized surface of constant energy $E<0$ of the Kepler problem. In this case, the particle moves along the circle $S^1$ in the fibre of the bundle \eqref{9.1}. Moser considered the redefinition of momenta and time and defined a mapping from the phase space $T^*\R^3$ into the cotangent bundle $T^*S^3\cong S^3\times\R^3$. He also showed that the surface of constant energy is a spherical subbundle $T^{1,1}=S(T^*S^3)$ in $T^*S^3$ and arrived at the same manifold as written out in \eqref{9.1}.

\smallskip

\noindent {\bf Hydrogen atom.}  The quantization in the KS approach was described in \cite{Chen1, Chen2}, and a description of all approaches to the Kepler problem, its quantization and references can be found in \cite{Cordany}. In the geometric quantization approach, the hydrogen atom is described by a fibre bundle $\CO(n-1, n-1)$ over $\CPP^1\times\CPP^1$ from \eqref{9.1} \cite{Hurt}. Sections of the complex line bundle $\CO(n-1, n-1)$ define the wave functions of the hydrogen atom in the coordinates of the KS approach. These wave functions correspond to the motion of a particle along a circle $S^1$ in spherical subbundles of the form
\begin{equation}\label{9.2}
\CO(0,0)\supset\CPP^1\times\CPP^1\times S^1\ \stackrel{S^1}{\longrightarrow}\ \CPP^1\times\CPP^1\ ,\ n=1\ ,
\end{equation}
\begin{equation}\label{9.3}
\CO(n-1, n-1)\supset  T^{1,1}/\Z_{n-1}   \ \stackrel{S^1}{\longrightarrow}\ \CPP^1\times\CPP^1\ ,\ n=2,3,...\ .
\end{equation}
The ground state \eqref{9.2} is purely quantum, since the particle motion occurs in fibres of the quantum bundle $\Lv$ over the regularized phase space $T^*S^3$. The states \eqref{9.3} are embedded into the surface of constant energy $T^{1,1}$ of the classical oscillator as subspaces. These subspaces $T^{1,1}/\Z_{n-1}$ are lens spaces similar to the lens spaces $S^3/\Z_n$ considered in this paper. The particle moves in them along a circle $S^1$ in the fibres of the bundle \eqref{9.3} similar to what was discussed in diagram \eqref{6.12}.

\bigskip

\noindent 
{\bf\large Acknowledgments}

\noindent
I am grateful to Tatiana Ivanova for useful remarks.


\bigskip

\begin{thebibliography}{99}
\bibitem{Dirac}
P.A.M. Dirac, {\it The principles of quantum mechanics}, Clarendon Press, Oxford, 1958.

\bibitem{Bar}
V.~Bargmann, ``On a Hilbert space of analytic functions and associated integral transform", Commun. Pure Appl. Math. {\bf 14} (1961) 187.

\bibitem{Segal}
I. Segal, {\it Mathematical problems of relativistic physics}, Chap. VI, in ``Proc. of the Summer Seminar, Boulder, Colorado, 1960, vol.II" (M.Kac, Ed.). 
Lectures in Applied Mathematics, AMS, Providence, Rhode Island, 1963.

\bibitem{Hall}
B.C.~Hall, ``Holomorphic methods in analysis and mathematical physics",\\ Contemp. Math. {\bf 260} (2000) 1.

\bibitem{Sni}
J. Sniatycki, {\it Geometric quantization and quantum mechanics}, \\Springer-Verlag, Berlin, 1980.

\bibitem{Wood}
N.M.J. Woodhouse, {\it Geometric quantization}, Clarendon Press, Oxford, 1980.

\bibitem{Hurt}
N.E. Hurt, 
{\it Geometric quantization in action}, \\ D.Reidel Publishing Company, Dordrecht, 1983.

\bibitem{Wells}
R.O. Wells, {\it Differential analysis on complex manifolds}, \\Springer-Verlag, New York, 1980.

\bibitem{Mumford}
D. Mumford , J. Fogarty and  F. Kirwan, {\it Geometric invariant theory},\\ Springer-Verlag, Berlin, 1994.

\bibitem{Thomas}
R.P. Thomas, ``Notes on GIT and symplectic reduction for bundles and varieties,"
Surveys in Differential Geometry, \textbf{10} (2006) 	
[arXiv:math/0512411 [math.AG]].

\bibitem{Hopf}
H.~Hopf, ``\"Uber die Abbildungen der dreidimensionalen Sph\"are auf die Kugelfl\"ache," \\
Math. Ann. {\bf 104} (1931) 637. 

\bibitem{Seifert}
H.~Seifert, ``Topologie dreidimensionaler gefaserter R\"aume,"\\ Acta Math. {\bf 60} (1933) 147.

\bibitem{SS}
D.D.~Sokolov and A.A.~Starobinsky, ``Structure of the curvature tensor on conical
singularities", Dokl. Akad. Nauk. USSR  {\bf 234} (1977) 1043.

\bibitem{BV}
M.~Barriola and A.~Vilenkin,
``Gravitational field of a global monopole'',\\
Phys. Rev. Lett. \textbf{63} (1989) 341.

\bibitem{Bordag}
M.~Bordag, K.~Kirsten and J.S.~Dowker,
``Heat kernels and functional determinants on the generalized cone'',
Commun. Math. Phys. \textbf{182} (1996) 371
[arXiv:hep-th/9602089 [hep-th]].

\bibitem{Popov1}
A.D.~Popov,
``Dirac particles, spin and photons,''
[arXiv:2508.21590 [hep-th]].

\bibitem{Popov2}
A.D.~Popov,
``Antiparticles in non-relativistic quantum mechanics,''\\
Mod. Math. Phys. {\bf 1} (2025)  4
[arXiv:2404.01756 [math-ph]].

\bibitem{Hooft}
G.~'t Hooft,
``The hidden ontological variable in quantum harmonic oscillators'',
Front. Quant. Sci. Tech. \textbf{3} (2024) 1505593
[arXiv:2407.18153 [quant-ph]].

\bibitem{S1}
J.R.~Hance and S.~Hossenfelder,
``What does it take to solve the measurement problem?'',
J. Phys. Comm. \textbf{6} (2022)  102001
[arXiv:2206.10445 [quant-ph]].

\bibitem{Tomaz}
A.A.~Tomaz, R.S.~Mattos and M.~Barbatti,
``The quantum measurement problem: a review of recent trends'', Philosophical Magazine C (2025) 1
[arXiv:2502.19278 [quant-ph]].

\bibitem{Stiefel}
P. Kustaanheimo and E. Stiefel,   ``Perturbation theory of Kepler motion based on spinor
regularization,"  J. Reine Angew Math. \textbf{218} (1965) 204. 

\bibitem{Moser}
J. Moser, ``Regularization of Kepler's problem and the averaging method on a manifold,"
Commun.  Pure and Appl. Math. {\bf23} (1970) 609.

\bibitem{Chen1}
A.C. Chen, ``Hydrogen atom as a four-dimensional oscillator,"
Phys. Rev. A {\bf 22} (1980) 333.

\bibitem{Chen2}
A.C. Chen and M. Kibler, ``Connection between the hydrogen atom and the four-dimensional oscillator"
Phys. Rev. A {\bf 31} (1985) 3960.

\bibitem{Cordany}
B. Cordany, {\it The Kepler problem: group theoretical aspect, regularization and quantization}, Birkh\"auser, Basel, 2013
\end{thebibliography}
\end{document}